\documentclass[aps,prd,preprint,endfloats,showpacs,superscriptaddress]{revtex4}
\usepackage{amsmath,amssymb,amsbsy,color}
\usepackage{moreverb}
\usepackage[dvips]{graphicx}
\usepackage{psfrag}
\usepackage{indentfirst}

\def\simgt{\rlap{\lower 3.5 pt\hbox{$\mathchar \sim$}}\raise 1pt \hbox
{$>$}}
\def\simlt{\rlap{\lower 3.5 pt\hbox{$\mathchar \sim$}}\raise 1pt \hbox
{$<$}}
\def\BE{\begin{equation}}
\def\EE{\end{equation}}
\def\BA{\begin{eqnarray}}
\def\EA{\end{eqnarray}}
\def\Bc{\begin{center}}
\def\Ec{\end{center}}
\def\LL{\left}
\def\RR{\right}

\def\nn{\nonumber\\}
\def\NN{\nonumber\\\nonumber\\}
\def\tr{\mbox{tr}\, }
\def\Tr{\mbox{Tr}\, }
\def\det{\mbox{det}}
\def\eps{\varepsilon}

\def\al{\alpha}
\def\bt{\beta}
\def\gm{\gamma}
\def\del{\delta}
\def\sig{\sigma}
\def\wtil{\widetilde}

\begin{document}

\draft
\preprint{UTHEP-506}
\preprint{UTCCS-P-12}

\title{Neutron electric dipole moment from lattice QCD}
\author{E.~Shintani}
\affiliation{
Graduate School of Pure and Applied Sciences, University of Tsukuba, 
Tsukuba, Ibaraki 305-8571, Japan }

\author{S.~Aoki}
\affiliation{
Graduate School of Pure and Applied Sciences, University of Tsukuba, 
Tsukuba, Ibaraki 305-8571, Japan }
\affiliation{
Riken BNL Research Center, Brookhaven National Laboratory, Upton, 11973,
USA}

\author{N.~Ishizuka}
\affiliation{
Graduate School of Pure and Applied Sciences, University of Tsukuba, 
Tsukuba, Ibaraki 305-8571, Japan }
\affiliation{
Center for Computational Sciences, University of Tsukuba, Tsukuba, 
Ibaraki 305-8577, Japan }

\author{K.~Kanaya}
\affiliation{
Graduate School of Pure and Applied Sciences, University of Tsukuba, 
Tsukuba, Ibaraki 305-8571, Japan }

\author{Y.~Kikukawa}
\affiliation{
Department of Physics, Nagoya University, Nagoya 464-8602, Japan }

\author{Y.~Kuramashi}
\affiliation{
Graduate School of Pure and Applied Sciences, University of Tsukuba, 
Tsukuba, Ibaraki 305-8571, Japan }
\affiliation{
Center for Computational Sciences, University of Tsukuba, Tsukuba, 
Ibaraki 305-8577, Japan }

\author{M.~Okawa}
\affiliation{
Department of Physics, Hiroshima University, Higashi-Hiroshima,
Hiroshima 739-8526, Japan }

\author{Y.~Taniguchi}
\affiliation{
Graduate School of Pure and Applied Sciences, University of Tsukuba, 
Tsukuba, Ibaraki 305-8571, Japan }
\affiliation{
Center for Computational Sciences, University of Tsukuba, Tsukuba, 
Ibaraki 305-8577, Japan }

\author{A.~Ukawa}
\affiliation{
Graduate School of Pure and Applied Sciences, University of Tsukuba, 
Tsukuba, Ibaraki 305-8571, Japan }
\affiliation{
Center for Computational Sciences, University of Tsukuba, Tsukuba, 
Ibaraki 305-8577, Japan }

\author{T.~Yoshi\'e}
\affiliation{
Graduate School of Pure and Applied Sciences, University of Tsukuba, 
Tsukuba, Ibaraki 305-8571, Japan }
\affiliation{
Center for Computational Sciences, University of Tsukuba, Tsukuba, 
Ibaraki 305-8577, Japan }

\collaboration{ CP-PACS Collaboration }

\date{\today}

\begin{abstract}
We carry out a feasibility study for the lattice QCD calculation of the
neutron
electric dipole moment (NEDM) in the presence of  the $\theta$ term.
We develop the strategy to obtain the nucleon EDM from the CP-odd 
electromagnetic form factor $F_3$ at small $\theta$, in which 
NEDM is given by $ \lim_{q^2\rightarrow 0}\theta F_3(q^2)/(2m_N) $ 
where $q$ is the momentum transfer and $m_N$ is the nucleon mass.
We first derive a formula which relates $F_3$, a matrix element of
the electromagnetic current between nucleon states, 
with vacuum expectation values of nucleons and/or the current.
In the expansion of $\theta$, 
the parity-odd part of the nucleon-current-nucleon three-point function
contains contributions not only from the parity-odd form factors but
also from
the parity-even form factors multiplied by the parity-odd part of the
nucleon two-point function,
and therefore the latter contribution must be subtracted to extract
$F_3$.
We then perform an explicit lattice calculation
employing the domain-wall quark action with the RG improved gauge action
in quenched QCD at $a^{-1}\simeq 2$ GeV on a $16^3\times 32\times 16$
lattice.
At the quark mass $m_f a =0.03$, corresponding to $m_\pi/m_\rho \simeq
0.63$,
we accumulate 730 configurations, which allow us to extract 
the parity-odd part in both two- and three-point functions.
Employing two different Dirac $\gamma$ matrix projections, we show that 
a consistent value for $F_3$ cannot be obtained without the subtraction 
described above. 
We obtain 
$F_3(q^2\simeq 0.58\,\textrm{GeV}^2)/(2m_N) =$
$-$0.024(5) $e\cdot$fm for the neutron and 
$F_3(q^2\simeq 0.58\,\textrm{GeV}^2)/(2m_N) =$
0.021(6) $e\cdot$fm for the proton. 

\end{abstract}


\maketitle

\section{Introduction}

In the strong interaction, one of the most 
stringent constraints on possible violation 
of parity (P) and time-reversal (T) symmetry  comes from 
the measurement of the electric dipole moment (EDM) for neutron (NEDM)
and 
proton (PEDM) ${\vec d}_{n,p}$.  
The current upper bound is given by
\BA
  |{\vec d}_n| < 6.3\times 10^{-26}\,\,e\cdot\textrm{cm}\,\textrm{(90\%
C.L.)}\,
\EA
for neutron\cite{Harris},  and 
\BE
  |{\vec d}_p| < 5.4 \times 10^{-24}\,\,e\cdot\textrm{cm}
\EE
for proton\cite{Dmitriev}, 
which are estimated from the results for EDM of mercury atom 
${}^{199}\rm{H}_{\rm{g}}$ given by 
$d({}^{199}\rm{H}_{\rm{g}})<2.1\times 10^{-28}\,e\cdot
\rm{cm}\,\textrm{(95\% C.L.)}$
\cite{Romalis}

On the other hand, QCD, which is regarded as the fundamental theory of
the
strong interaction, allows a gauge invariant, renormalizable CP odd
operator
in the Lagrangian, called the $\theta$ term:  
\BA
&&
i\frac{\theta}{32\pi^2}\int d^4x\,\wtil G_{\mu\nu}(x)G_{\mu\nu}(x)
   \label{L_QCD}, \qquad
   \wtil G_{\mu\nu}(x) = \frac{1}{4}\eps_{\mu\nu\al\bt}G_{\al\bt}(x)
\EA
in Euclidean space-time
with $G_{\mu\nu}$ the field strength of gluon.
Some model estimations\cite{Crewther,Vecchia}
yield
\BE
|{\vec d}_n| \sim \theta \times O(10^{-15}\sim 10^{-16}) \,e\cdot{\rm
cm}, 
\EE
and this leads to a bound $\theta \simlt O(10^{-10})$.
Hence $\theta$ must be small or even must vanish in QCD.

A smallness of $\theta$ in the QCD sector, however, 
is not protected in the presence of the
electroweak sector of the standard model, where the quark mass matrix, 
arising from Yukawa couplings to the Higgs field, 
may be written as
\BE
  \bar\psi^R_i(x) M_{ij}\psi^L_j(x) + \bar\psi^L_i(x)
M^{\dag}_{ij}\psi^R_j(x),
\EE
where $\psi_L$ and $\psi_R$ represent left  and right handed quark
fields
with flavor indices $i,j$.  Diagonalizing the mass matrix and 
making it real, the parameter $\theta$ becomes
\BE
  \theta = \theta_{\rm QCD} + \arg\det M ,
\EE
where $\theta_{\rm QCD}$ is the original $\theta$ parameter in QCD.
Therefore, either $\theta_{\rm QCD}$ and $\arg\det M$ are individually
small, 
or the two contributions cancel out to the degree that the experimental
upper bound on NEDM is satisfied.
In either of the two cases, it seems necessary to explain why Nature
chooses 
such a small value for $\theta$; this is the ``strong CP problem''.
One of the most attractive explanations proposed so far is the 
Peccei-Quinn mechanism\cite{Peccei}.  Unfortunately, 
the axion, a new particle predicted by this mechanism, 
has not been experimentally observed yet.

The present model estimations of the hadronic contribution to
NEDM, $|{\vec d}_n|/\theta$, are maybe enough to convince 
us of the the smallness of $\theta$\cite{QCDsum,ChPT}.
However, a theoretically reliable and accurate estimation for NEDM
will be required to determine the value of $\theta$, 
if non-zero value of  NEDM
is observed in future experiments.
Lattice QCD calculations seem ideal for this requirement.
Indeed more than 15 years
ago the first attempt was made to estimate  NEDM in a quenched
lattice QCD simulation\cite{Aoki1} by calculating the spin dependent
energy
difference of the nucleon
in the presence of the uniform static electric field.
In this simulation, the $\theta$ was converted to the phase of the 
quark mass term by the chiral rotation, and a non-zero value of  NEDM
was 
obtained for non-zero $\theta$.  Unfortunately this non-zero value 
of NEDM turned out\cite{Aoki2} to be a lattice artifact
due to the explicit chiral symmetry breaking of the Wilson fermion
action 
employed in the simulation; if the $\theta$ term appears in the quark
mass,
an additional disconnected contribution must be included in the quenched
calculation of Ref.\cite{Aoki1}.

After this attempt, there has been no further lattice investigation
on this problem for a long time.
Recently, a new lattice strategy \cite{Guadagnoli} for the
extraction of NEDM has been proposed, and
a lattice related investigation on NEDM has been made\cite{Faccioli}.
Also, a preliminary lattice QCD result of NEDM estimated from the
nucleon 
electromagnetic three-point function has been reported
at the lattice 2004 conference\cite{Blum}.
The value of NEDM divided by $\theta$ is consistent with zero within 
statistical errors in this calculation, however.
In all attempts, an expansion in $\theta$ have been used to avoid
the complex action due to the  $\theta$ term.

Since chiral symmetry seems important in the calculation of
NEDM\cite{Aoki3},
it may be preferable to use lattice fermion formulations having good
chiral 
symmetry, such as the domain-wall fermion\cite{DW} or the overlap 
fermion\cite{overlap}.
In these formulations, the bosonic definitions of the topological 
charge agree well with the fermionic definitions.
In fact the domain-wall fermion was employed in Ref.\cite{Blum}.

In this paper we propose a new method to calculate NEDM in lattice QCD. 
We consider the strategy of extracting NEDM from the CP-odd part 
of the electromagnetic form factor of the nucleon, as in the 
recent lattice calculation \cite{Blum} as well as in some past 
model calculations \cite{QCDsum,ChPT}.
We first examine the $\theta$ dependence of the three-point 
(two nucleons and the electromagnetic current) function
by expanding up to
first order in $\theta$ under the assumption that
$\theta$ has fairly small value.
An important finding is that the 
parity-odd part of the three-point function contains
a contribution coming from the parity-even form factors 
multiplied by the parity-odd part of the nucleon two-point function.
This 
contribution has to be subtracted from the three-point function  
in order to properly extract the CP-odd form factor relevant for NEDM. 
This point is not considered in Ref.\cite{Blum}.
We present the formula to extract the relevant form factor in 
terms of the nucleon two- and three-point functions,
which is valid not only for lattice simulations but also for 
general cases as long as chiral symmetry of quarks is well formulated
in the calculations.

We test our formula in a lattice QCD simulation 
using the domain-wall fermion in the quenched approximation. 
A renormalization group (RG) improved gauge action is employed
based on our experience that 
the residual chiral symmetry breaking due to a finite fifth dimensional 
extent is smaller than in the case for the plaquette gauge action 
at a similar lattice spacing\cite{CP-PACS1}. 
Since our primary purpose is to check the feasibility of the
lattice QCD calculation of NEDM based on our formula,
we focus on accumulating the statistics to obtain a sufficient
sampling of topological charge
on a $16^3\times 32\times 16$ lattice at $a^{-1}\simeq 2$ GeV 
with the quark mass of $m_f=0.03$, corresponding to 
$m_\pi/m_\rho \simeq 0.63$.
Our numerical results reveal that the contribution 
to the three-point function from
the parity-even form factors 
multiplied by the parity-odd part of the nucleon two-point function
is really significant.
We demonstrate the correctness of our formula by
showing that two independent ways to extract the relevant form factor
give consistent results only if the contribution from the parity-even
form factors is properly subtracted. 
 
This paper is organized as follows.
In Sec.~\ref{sec: EDM_formula} 
we derive the formula to extract the relevant form factor 
in terms of the two- and three-point functions 
to first order in the expansion in $\theta$. 
Section~\ref{sec:lattice_formulation} contains simulation parameters
and technical details.  
In Sec.~\ref{sec:result} we present the simulation results 
for the CP-odd form factors together with the CP-even ones.
It is demonstrated that we cannot obtain the correct value 
for the CP-odd form factor without subtracting the 
contribution from the parity-even form factors.
We conclude our investigation in Sec.~\ref{sec:conclusion}.

\section{Nucleon electric dipole moment from correlation functions
\label{sec: EDM_formula}}
\subsection{Form factors and EDM}
\label{subsec:EDM_def}
We consider the electromagnetic form factors of the nucleon, defined by
\BA
\langle N (\vec p,s)\vert J_\mu^{\rm EM}\vert
N(\vec{p^\prime},s^\prime)\rangle 
&=& \bar u(\vec p,s)\left[ \frac{F_3(q^2)}{2m_N} q_\nu
\sigma_{\mu\nu}\gm_5+\cdots
\right] u(\vec{p^\prime},s^\prime),
\EA
where $q=p-p^\prime$ is the momentum transfer, $\vert N(\vec
p,s)\rangle$ is 
the on-shell nucleon state with momentum $\vec{p}$, energy
$p_0=\sqrt{m_N^2+\vec{p}^2}$ 
and helicity $s$. 
The electromagnetic current for quarks, $J_\mu^{\rm EM} $, is given by
\BE
J_\mu^{\rm EM} = \sum_f e_f\bar\psi_f  \gamma_\mu \psi_f, 
\EE
where $\psi_f $ is the quark field with flavor $f$ and the electric
charge $e_f$.

In the small momentum transfer limit, $q\longrightarrow 0$, 
the above form factor is
described by the following effective interaction:
\BE
\bar N (x) \left[\frac{F_3(0)}{2m_N} \sigma_{\mu\nu}\gm_5 \partial_\nu
A_\mu (x) + \cdots
\right] N(x),
\EE
where $A_\mu (x)$ is the U(1) electromagnetic field. We can define
the electric dipole moment ${\vec d}_N$ as
\BE
\vert {\vec d}_N\vert = \lim_{q^2\rightarrow
0}\frac{F_3(q^2)}{2m_N}=\frac{F_3(0)}{2m_N}  .
\EE 
So far we do not specify CP properties of the system and therefore
$\theta$ dependences are omitted in the above expressions.

\subsection{Extraction of form factors from correlation functions}
In the lattice calculation we must extract the form factors from the
three-point correlation functions such as
\BA
G_{NJ_\mu N}^\theta (q, t, \tau ) &\equiv &
\langle \theta \vert N(\vec p,t) J_{\mu}^{\rm EM}(\vec q,\tau)\bar
N(\vec{p^\prime},0)\vert 
\theta \rangle,
\EA 
where $N(\vec p, t)$ and $\bar N(\vec p, t)$ are the 
interpolating fields of the nucleon at time $t$,
which contain annihilation and creation operators with momentum
$\vec{p}$.
Here $\vert \theta \rangle $ is the $\theta$ vacuum, which may be
defined as
\BE
\vert \theta \rangle = e^{i \theta Q_5 } \vert 0 \rangle, \quad
Q_5 =\int d^3 x K_0(\vec{x}, t) ,
\EE
using the conserved (but gauge non-invariant) chiral current $K_\mu$. 
Throughout this paper,
we do not use the explicit representation of $\vert \theta \rangle$.
Inserting the complete set of states between the interpolating fields
and 
the current, we obtain
\BA
G_{NJ_\mu N}^\theta (q, t, \tau ) &=&
e^{-E_{N^\theta}(t-\tau)}e^{-E_{N^\theta}^\prime t}\nn
&\times&
\sum_{s,s^\prime}\langle \theta \vert N \vert N^\theta
(\vec{p},s)\rangle
\langle N^\theta (\vec{p},s)\vert J_\mu^{\rm EM}\vert N^\theta
(\vec{p^\prime},s^\prime)\rangle
\langle N^\theta (\vec{p^\prime},s^\prime)\vert \bar N\vert\theta\rangle
+ \cdots,
\EA 
where $E_{N^\theta} =\sqrt{\vec{p}^2+m_{N^\theta}^2}$,
$E_{N^\theta}^\prime =\sqrt{\vec{p^\prime}^2+m_{N^\theta}^2}$, and
the dots represent exponentially suppressed contributions,
which are assumed to be negligible.
Hereafter we represent explicitly superscript of $\theta$ on $\theta$
vacuum effects.
Since the interpolating fields $N$ and $\bar N$ can be expanded as
\BA
N(\vec{p},0) &=& Z_N^\theta \sum_s  a_{N^\theta}(\vec{p},s) u_N^\theta
(\vec p,s)
+\cdots, \nn
\bar N(\vec{p},0) &=& (Z_N^\theta)^* \sum_s 
a_{N^\theta}^\dagger(\vec{p},s) \bar 
u_N^\theta (\vec p,s)+\cdots ,
\label{eq:NfieldExpand}
\EA
we have
\BA
\langle \theta \vert N \vert N^\theta (\vec{p},s)\rangle &=& Z_N^\theta
u_N^\theta (\vec p,s), \nn
\langle N^\theta (\vec{p^\prime},s^\prime)\vert \bar N\vert\theta\rangle
&=&
(Z_N^\theta)^* \bar u_N^\theta (\vec{ p^\prime},s^\prime) ,
\EA
since 
\BA
\vert N^\theta (\vec{p},s)\rangle &=& a_{N^\theta}^\dagger(\vec{p},s)
\vert\theta\rangle \\
\langle N^\theta (\vec{p},s)\vert &=&
\langle \theta \vert a_{N^\theta}(\vec{p},s) .
\EA
In eq.~(\ref{eq:NfieldExpand}) 
the dots represent contributions from all other possible states 
including the negative energy state of the nucleon, the negative parity 
states or other excited states. 
Except a special case considered in the appendix, we will not
consider these states in this paper.
The spinors $u_N^\theta (\vec p,s)$ and $\bar u_N^\theta (\vec p,s)$ are
on-shell nucleon 
wave functions which satisfy the Dirac equation:
\BA
\left(i\gamma\cdot p + m_{N^{\theta}}e^{-i f_N(\theta)\gamma_5} \right)
u_N^\theta(\vec{p},s)&=&
\bar u_N^\theta(\vec{p},s)\left(i\gamma\cdot p + m_{N^{\theta}}e^{-i
f_N(\theta)\gamma_5} \right) =0 .
\EA
Since CP is broken in the $\theta$ vacuum, a CP non-invariant phase
factor 
$e^{if_N(\theta)\gamma_5}$ can appear in the mass term. This means that
$f_N(\theta)$ must be odd in $\theta$, while $m_N$ and $Z_N$ are even in
$\theta$.
Therefore we have
\BA
m_{N^\theta} &=& m_N + O(\theta^2), \quad Z_N^\theta = Z_N + O(\theta^2)
\nn
f_N(\theta) &=& f_N^1 \theta + O(\theta^3) 
\EA
for small $\theta$.
The bi-spinor projection satisfies
\BA
\sum_{s} u_N^\theta (\vec{p},s)\bar u_N^\theta (\vec{p},s)
&=& \frac{-i\gamma\cdot p + m_{N^\theta}e^{i
f_N(\theta)\gamma_5}}{2E_{N^\theta}}
\label{eq:2-pt_in_theta}
\EA
with the normalization that $\bar u_N^\theta (\vec{p},s)
u_N^\theta(\vec{p},s^\prime) 
= \displaystyle \frac{m_{N^{\theta}}\cos
f_N(\theta)}{E_{N^{\theta}}}\delta_{s,s'}$.

The form factor,  which we wish to extract from the three-point
function, 
is written as
\BA
\langle N^\theta (\vec{p},s)\vert J_\mu^{\rm EM}\vert N^\theta
(\vec{p^\prime},s^\prime)\rangle
&=& \bar u_N^\theta (\vec{p},s) W_\mu^\theta (q) u_N^\theta
(\vec{p^\prime},s^\prime)
\EA
It satisfies
\BE
\bar u_N^\theta (\vec{p},s)q_\mu W_\mu^\theta (q) u_N^\theta
(\vec{p^\prime},s^\prime)
=0
\EE
from the current conservation $\partial_\mu J_\mu^{\rm EM} = 0$.
Taking account of the parity properties,
the form factors in general should have the form that
\BA
W_\mu^\theta (q) &=& g(\theta^2)\  W_\mu^{\rm even}(q)
+i\theta\ h(\theta^2)\ W_\mu^{\rm odd}(q), 
\EA
where we take $g(0)=h(0)=1$ for their normalizations and 
\BA
W_\mu^{\rm even}(q)&=& \gamma_\mu
F_1(q^2)+\frac{F_2(q^2)}{2m_N}q_\nu\sigma_{\mu\nu},\\
W_\mu^{\rm odd}(q)&=& \frac{F_3(q^2)}{2m_N}q_\nu\sigma_{\mu\nu}\gamma_5
+ F_A(q^2)\left(q_\mu \gamma\cdot q-\gamma_\mu q^2\right)\gamma_5 
\EA
with  $\sig_{\mu\nu}=i[\gm_{\mu},\gm_{\nu}]/2$.
Here $F_{1,2}(q^2)$ are the electromagnetic form factors 
and $F_A(q^2)$ is called ``anapole form factor''.
As mentioned in Sec.~\ref{subsec:EDM_def}, $F_3(0)/(2m_N)$ gives the
nucleon 
electric dipole moment. 
For small $\theta$, we can expand $g(\theta^2) = 1 + O(\theta^2)$
and $h(\theta^2)=1 + O(\theta^2)$.

Using all\ the necessary expressions and expanding 
in terms of $\theta$, we obtain
\BA
G_{NJ_\mu N}^\theta (q, t, \tau ) &=& \vert Z_N\vert^2
e^{-E_{N}(t-\tau)}e^{-E_{N}^\prime t}
\frac{-i\gamma\cdot p + m_{N}(1+i f_N^1\theta \gamma_5)}{2E_{N}}
\nn
&\times& 
\left[W_\mu^{\rm even}(q) +i\theta W_\mu^{\rm odd}(q) \right]
\frac{-i\gamma\cdot p^\prime + m_{N}(1+i f_N^1\theta
\gamma_5)}{2E_{N}^\prime}
+O(\theta^2).
\label{eq:3pt_ff}
\EA
On the other hand, $G_{NJ_\mu N}^\theta (q, t, \tau )$ can be evaluated
by the path integral as
\BA
 G_{NJ_\mu N}^\theta (q, t, \tau ) &=& 
 \langle N(\vec{p},t) J_\mu^{\rm EM}(\vec q,\tau) \bar
N(\vec{p^\prime},0)e^{i\theta Q}\rangle
 \nn
 &=& G_{NJ_\mu N} (q, t, \tau ) + i\theta G_{NJ_\mu N}^Q (q, t, \tau ) +
O(\theta^2)
\label{eq:3pt_pi}
\EA
with
\BA
G_{NJ_\mu N} (q, t, \tau )&=&\langle \vec N(\vec{p},t) J_\mu^{\rm
EM}(\vec q,\tau) \bar
 N(\vec{p^\prime},0)\rangle, \\
G_{NJ_\mu N}^Q (q, t, \tau )&=&\langle \vec N(\vec{p},t) J_\mu^{\rm
EM}(\vec q,\tau) 
\bar N(\vec{p^\prime},0) Q\rangle,
\EA
where $N$, $\bar N$, and $J_\mu^{\rm EM}$ are c-number operators in the
path-integral, 
albeit using the same notation as the quantum operators, 
and $Q$ is the topological charge, defined by
\BE
Q=\frac{1}{32\pi^2}\int d^4x\,\wtil G_{\mu\nu}(x)G_{\mu\nu}(x)
\EE
in the continuum theory.  
The symbol $\langle {\cal O}\rangle$ denotes the path-integral average
of 
an operator $\cal O$
in QCD with $\theta=0$, which is given by
\BE
\langle {\cal O} \rangle = \int {\cal D}\psi {\cal D}\bar \psi {\cal D}
A_\mu
\,  e^{S_G(A) + \bar\psi D(A)\psi }\, {\cal O},
\EE 
where $S_G(A)$ and $\bar\psi D(A)\psi$ are the QCD gauge and quark
actions,
respectively.

By comparing eqs.(\ref{eq:3pt_ff}) and (\ref{eq:3pt_pi}) we obtain
\BA
G_{NJ_\mu N} (q, t, \tau ) &=& \vert Z_N\vert^2
e^{-E_{N}(t-\tau)}e^{-E_{N}^\prime t}
\frac{-i\gamma\cdot p + m_{N}}{2E_{N}} W_\mu^{\rm even}(q)
\frac{-i\gamma\cdot p^\prime + m_{N}}{2E_{N}^\prime},
\label{eq:3pt_even}\\
G_{NJ_\mu N}^Q (q, t, \tau ) &=&  \vert Z_N\vert^2
e^{-E_{N}(t-\tau)}e^{-E_{N}^\prime t}
\Bigl[
\frac{-i\gamma\cdot p + m_{N}}{2E_{N}} W_\mu^{\rm odd}(q)
\frac{-i\gamma\cdot p^\prime + m_{N}}{2E_{N}^\prime} \nn 
&+&
\frac{f_N^1m_N}{2E_N}\gamma_5 W_\mu^{\rm even}(q)
\frac{-i\gamma\cdot p^\prime + m_{N}}{2E_{N}^\prime} \nn
&+& \frac{-i\gamma\cdot p + m_{N}}{2E_{N}} W_\mu^{\rm even}(q)
\frac{f_N^1m_N}{2E_N^\prime}\gamma_5
\Bigr].\label{eq:3pt_odd}
\EA
These formulae are one of the main results of this paper.  They 
show that the parity-even form factor $W_\mu^{\rm even}$
multiplied with the parity-odd term $f_N^1$ of the two-point function 
has to be removed from the three-point function $G_{NJ_\mu J}^Q$ 
at order $\theta$, in order to obtain $W_\mu^{\rm odd}(q)$.
We will see in Sec.~\ref{sec:result} how significant 
this subtraction is.

\subsection{Extraction of $f_N^1$ from the nucleon two-point function}
\label{sec:formula_2pt}
In order to determine $f_N^1$ we consider the two-point function of the
nucleon given by
\BA
G_{NN}^\theta (p,t) &\equiv & \langle \theta\vert N(\vec{p},t) \bar
N(\vec{p},0)\vert \theta\rangle \nn
&=& \vert Z_N^\theta\vert^2 e^{-E_{N^\theta} t}\frac{-i\gamma\cdot p +
m_{N^\theta}
e^{if_N(\theta)\gamma_5}}{2E_{N^\theta}} 
\label{eq:2ptOperator}\\
&=& \langle N(\vec{p},t) \bar N(\vec{p},0) e^{i\theta Q}\rangle  .
\label{eq:2ptPathIntegral}
\EA 
By comparing the operator expression (\ref{eq:2ptOperator}) with the
path integral expression
(\ref{eq:2ptPathIntegral}) at each order of $\theta$, we obtain 
\BA
G_{NN}(\vec{p},t) &\equiv & \langle N(\vec{p},t) \bar
N(\vec{p},0)\rangle  =
\vert Z_N\vert^2 e^{-E_{N} t}\frac{-i\gamma\cdot p + m_{N}}{2E_{N}} 
\EA
at $\theta^0$ and
\BA
G_{NN}^Q(\vec{p},t) &\equiv & \langle N(\vec{p},t) \bar N(\vec{p},0)
Q\rangle  =
\vert Z_N\vert^2 e^{-E_{N} t}\frac{ f_N^1m_{N}}{2E_{N}} \gamma_5
\EA
at $\theta^1$.
These formulae tell us that $f_N^1$ can be determined numerically from 
the nucleon propagators through the appropriate insertion of $Q$.

\section{Simulation details}
\label{sec:lattice_formulation}
\subsection{Fermion and gauge actions}
For the quark field
we employ the domain-wall fermion action\cite{DW}:
\BA
  S_F &=& -\sum_{x,s,x',s'}\bar\psi(x,s)D_{\rm DW}(x,s;x',s')\psi(x',s')
+ \sum_x m_f\bar q(x)q(x),\\
  D_{\rm DW}(x,s;x's') &=& D^4(x,x')\del_{s,s'} + \del_{x,x'}D^5(s,s') +
(M-5)\del_{x,x'}\del_{s,s'},\\
  D^4(x,x') &=&
\sum_{\mu}\frac{1}{2}\LL[(1-\gm_{\mu})U_{\mu}(x)\del_{x+\hat\mu,x'}
            + (1+\gm_{\mu})U^{\dag}_{\mu}(x')\del_{x,x'+\hat\mu}\RR],\\
  D^5(s,s') &=& \LL\{
\begin{array}{cc}
  \frac{1}{2}(1-\gm_5)\del_{2,s'} & (s=1)\\
  \frac{1}{2}(1-\gm_5)\del_{s+1,s'} + \frac{1}{2}(1+\gm_5)\del_{s-1,s'}
& (1<s<N_s)\\
  \frac{1}{2}(1+\gm_5)\del_{N_s-1,s'} & (s=N_s)
\end{array}  , \RR.
\EA
where $x,x'$ are four dimensional space-time coordinates,  $s,s'$ are 
coordinates in the fifth
 dimension of length $N_s$, and $M$ is the domain-wall height. 
Here $q,\,\bar q$ are four dimensional physical quark fields given by
\BA
  q(x)      &=& \frac{1}{2}(1-\gm_5)\psi(x,1) +
\frac{1}{2}(1+\gm_5)\psi(x,N_s),\\
  \bar q(x) &=& \bar\psi(x,N_s)\frac{1}{2}(1-\gm_5) +
\bar\psi(x,1)\frac{1}{2}(1+\gm_5)
\EA
and $m_f$ is the bare quark mass. 
This action has exact chiral symmetry at $N_s\rightarrow\infty$,
and therefore it is expected that the fermion fields respond properly to
the topological 
charge of the gauge fields.

For the gauge field we employ the following improved action:
\BE
  S_G = \frac{\bt}{6}\LL\{c_0\sum_{\rm plaq}\Tr U_{\rm plaq} +
c_1\sum_{\rm rect}\Tr U_{\rm rect}\RR\},
  \label{eq:Gauge_action}
\EE
where the first term represents the plaquette and the second the
$1\times 2$ 
rectangle loop. In this paper we take $c_0=3.648$ and $c_1=-0.331$,
proposed by Iwasaki using the renormalization group
analysis\cite{iwasaki}.
The action with this parameter choice is called 
the RG improved  gauge action.

\subsection{Topological charge}
In continuum QCD, the topological charge density is defined by
\BE
  q(x) =
\frac{1}{32\pi^2}\eps_{\mu\nu\al\bt}\Tr\,G_{\mu\nu}(x)G_{\al\bt}(x)
\EE
and the topological charge is given by the integration of $q(x)$ over
space-time:
\BE
  Q = \int d^4x\,q(x).
\EE

\begin{figure}[h]
\Bc
\psfrag{x}{ }
\includegraphics[width=25mm, angle=0]{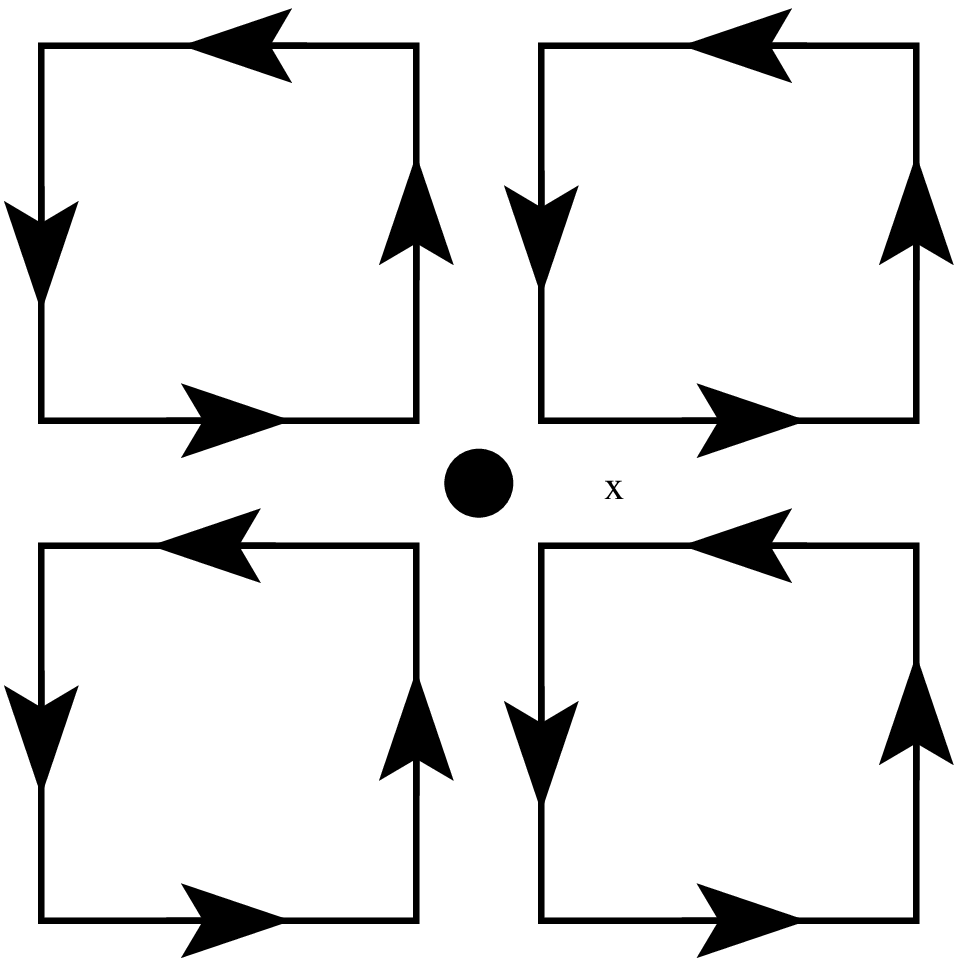}
\hskip 20mm
\includegraphics[width=25mm, angle=0]{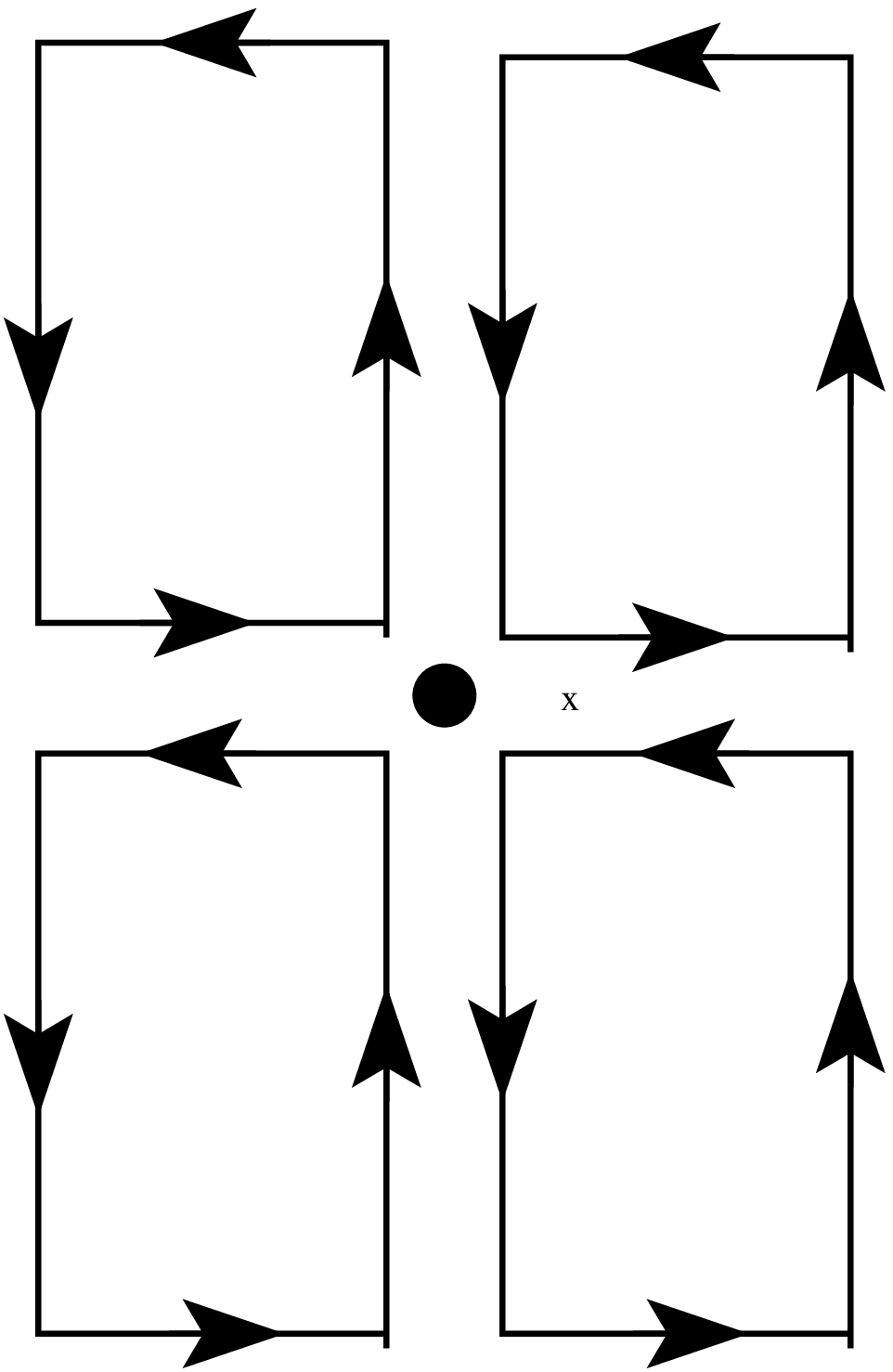}
\hskip 10mm
\includegraphics[width=40mm, angle=0]{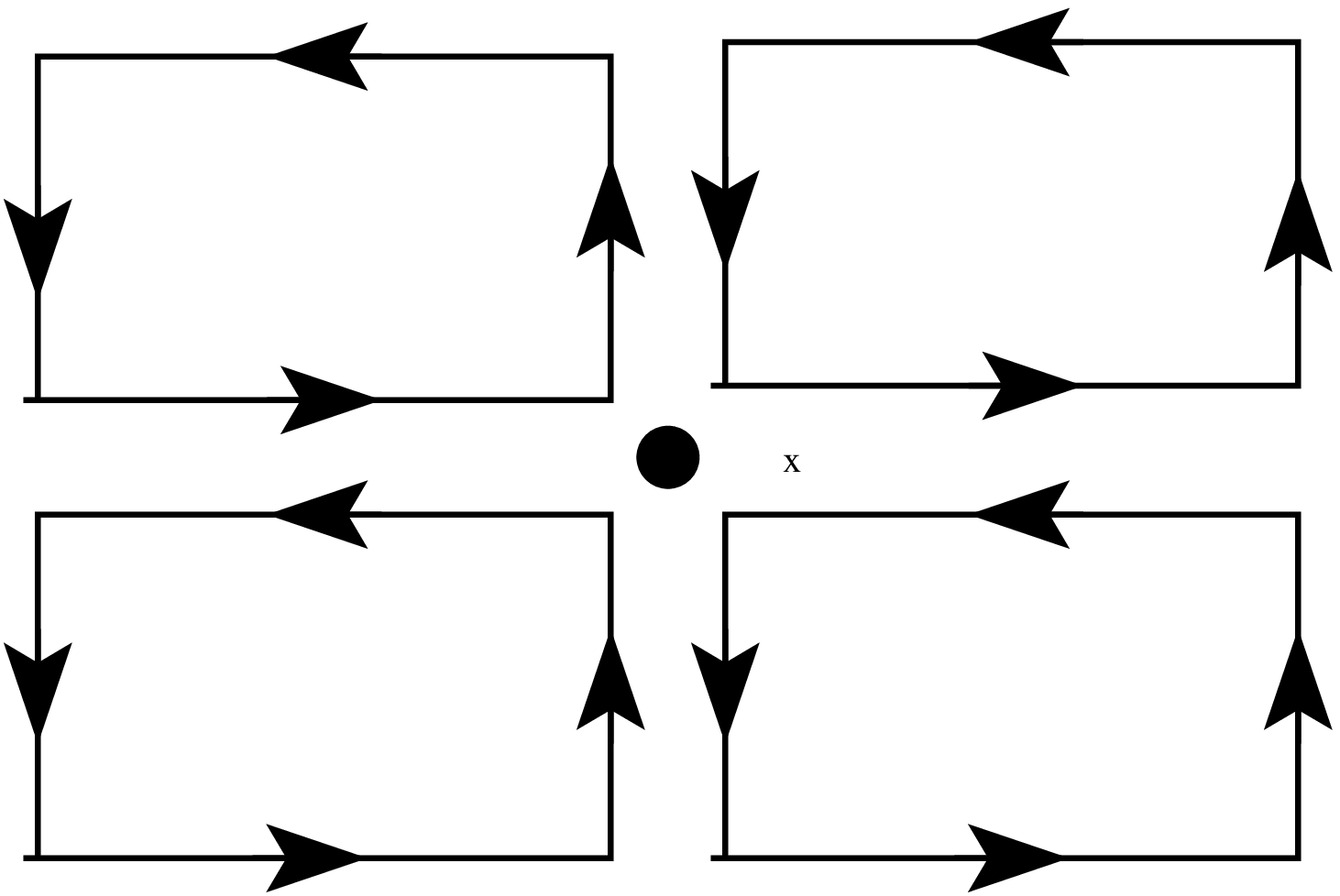}
\Ec
\caption{Plaquette clover (left) and rectangular clovers (middle and
right).}
\label{fig:clover}
\end{figure}

In our lattice calculation we define the topological charge by the
improved bosonic 
form\cite{CP-PACS2}, which is given by 
\BA
  Q_{\rm imp} &=& \sum_x\LL[c_0 Q_{\rm plaq}(x) + c_1 Q_{\rm
rect}(x)\RR] \label{top_def}\\
  Q_{\rm plaq} &=&
\frac{1}{32\pi^2}\eps_{\mu\nu\al\bt}\Tr(F^P_{\mu\nu}F^P_{\al\bt}),\quad
  Q_{\rm rect} =
\frac{2}{32\pi^2}\eps_{\mu\nu\al\bt}\Tr(F^R_{\mu\nu}F^R_{\al\bt})
\EA
where $F_{\mu\nu}^P$ and $F_{\mu\nu}^R$ are field strengths constructed
from
the plaquette clover and the rectangular clovers as
depicted in Fig.~\ref{fig:clover}.  Their explicit expressions are given
by
\BA
F^P_{\mu\nu}(x)&=&\frac{\rm Im}{4}\sum_{x,\mu,\nu}\tr\Bigl[
U_{\mu}(x)U_{\nu}(x+\hat\mu)U_{\mu}^{\dag}(x+\hat\nu)U_{\nu}^{\dag}(x)\nn
&+&
U_{\nu}(x)U^{\dag}_{\mu}(x-\hat\mu+\hat\nu)U_{\nu}^{\dag}(x-\hat\mu)U_{\mu}(x-\hat\mu)\nn
&+& U^{\dag}_{\mu}(x-\hat\mu)U^{\dag}_{\nu}(x-\hat\mu-\hat\nu)
    U_{\mu}(x-\hat\mu-\hat\nu)U_{\nu}(x-\hat\nu)\nn
&+&
U^{\dag}_{\nu}(x-\hat\nu)U_{\mu}(x-\hat\nu)U_{\nu}(x+\hat\mu-\hat\nu)U^{\dag}_{\mu}(x)
\Bigr], \\
F^R_{\mu\nu}(x) &=& \frac{\rm Im}{8}\sum_{x,\mu,\nu}\biggl\{\tr\Big[
    U_{\mu}(x)U_{\nu}(x+\hat\mu)U_{\nu}(x+\hat\mu+\hat\nu)U_{\mu}^{\dag}(x+2\hat\nu)
    U_{\nu}^{\dag}(x+\hat\nu)U_{\nu}^{\dag}(x)\nn
&+&
U_{\nu}(x)U_{\nu}(x+\hat\nu)U^{\dag}_{\mu}(x-\hat\mu+2\hat\nu)U_{\nu}^{\dag}(x-\hat\mu+\hat\nu)
    U_{\nu}^{\dag}(x-\hat\mu)U_{\mu}(x-\hat\mu)\nn
&+&
U^{\dag}_{\mu}(x-\hat\mu)U^{\dag}_{\nu}(x-\hat\mu-\hat\nu)U^{\dag}_{\nu}(x-\hat\mu-2\hat\nu)
    U_{\mu}(x-\hat\mu-2\hat\nu)U_{\nu}(x-2\hat\nu)U_{\nu}(x-\hat\nu)\nn
&+&
U^{\dag}_{\nu}(x-\hat\nu)U^{\dag}_{\nu}(x-2\hat\nu)U_{\mu}(x-2\hat\nu)
U_{\nu}(x+\hat\mu-2\hat\nu)U_{\nu}(x+\hat\mu-\hat\nu)U^{\dag}_{\mu}(x)\Big]\nn
&+& \tr\Big[ 
    U_{\mu}(x)U_{\mu}(x+\hat\mu)U_{\nu}(x+2\hat\mu)U_{\mu}^{\dag}(x+\hat\nu+\hat\mu)
    U_{\mu}^{\dag}(x+\hat\nu)U_{\nu}^{\dag}(x)\nn
&+&
U_{\nu}(x)U^{\dag}_{\mu}(x-\hat\mu+\hat\nu)U_{\mu}^{\dag}(x-2\hat\mu+\hat\nu)
    U_{\nu}^{\dag}(x-2\hat\mu)U_{\mu}(x-2\hat\mu)U_{\mu}(x-\hat\mu)\nn
&+&
U^{\dag}_{\mu}(x-\hat\mu)U^{\dag}_{\mu}(x-2\hat\mu)U^{\dag}_{\nu}(x-2\hat\mu-\hat\nu)
U_{\mu}(x-2\hat\mu-\hat\nu)U_{\mu}(x-\hat\mu-\hat\nu)U_{\nu}(x-\hat\nu)\nn
&+&
U^{\dag}_{\nu}(x-\hat\nu)U_{\mu}(x-\hat\nu)U_{\mu}(x-\hat\nu+\hat\mu)U_{\nu}(x+2\hat\mu-\hat\nu)
    U^{\dag}_{\mu}(x+\hat\mu)U^{\dag}_{\mu}(x)\Big]
\biggr\}.
\EA
In order to remove $O(a^2)$ discretization errors in $Q_{\rm imp}$ at 
the tree level,
we take $c_1=-1/12$ and $c_0=1-8c_1=5/3$\cite{impQ}.


\subsection{Nucleon three-point functions and form factors}

\begin{figure}
\Bc
\psfrag{p2}{ }
\psfrag{p1}{ }\psfrag{sink}{ }\psfrag{source}{ }\psfrag{operator}{ }
\psfrag{p0}{ }
\psfrag{t1}{ }
\psfrag{tau}{ }
\psfrag{t2}{ }
\psfrag{1}{ }\psfrag{2}{ }
\psfrag{u}{\Large u}
\psfrag{d}{\Large d}
\psfrag{aa}{$(a)$}
\psfrag{bb}{$(b)$}
\psfrag{cc}{$(c)$}
\psfrag{dd}{$(d)$}
\psfrag{ee}{$(e)$}
\psfrag{ff}{$(f)$}
\vskip 11cm 
\includegraphics[height=60mm]{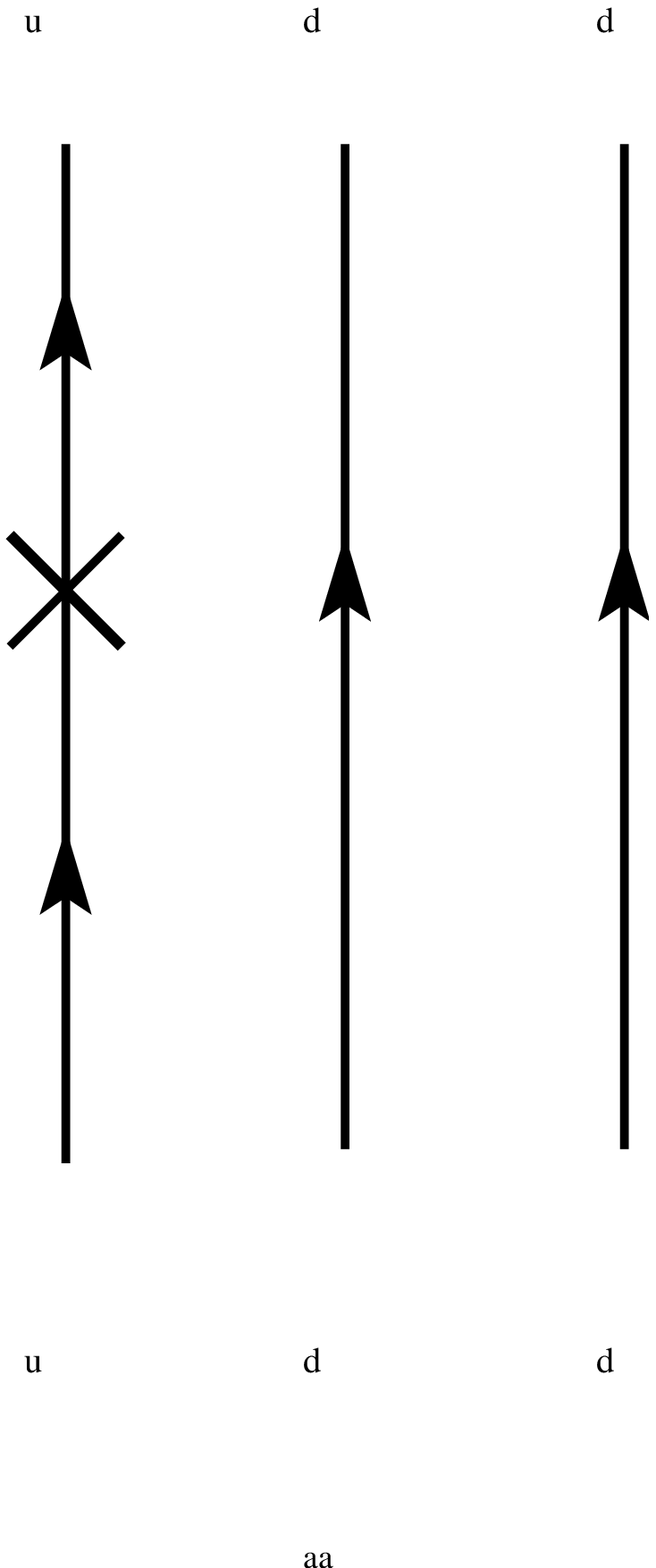}
\hspace{3cm}
\includegraphics[height=60mm]{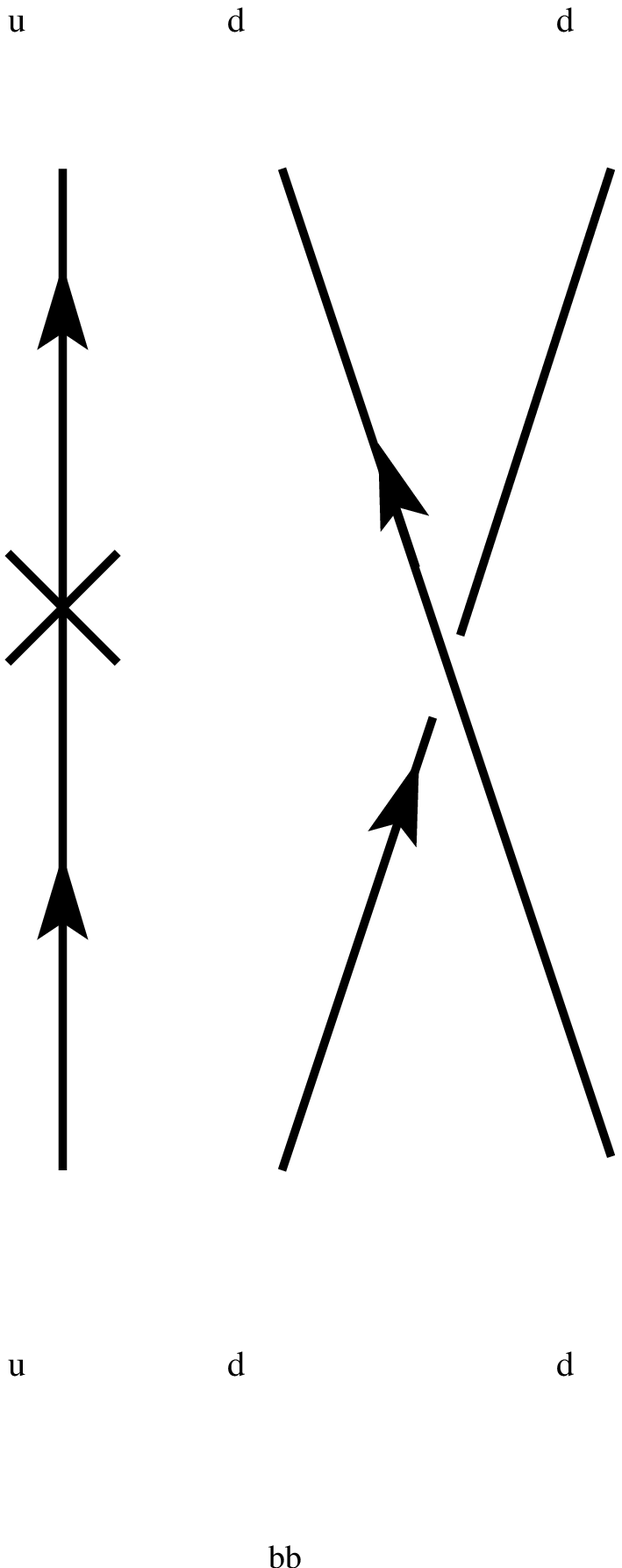}\\
\vspace{1.2cm}
\hspace{-.1cm}
\includegraphics[height=60mm]{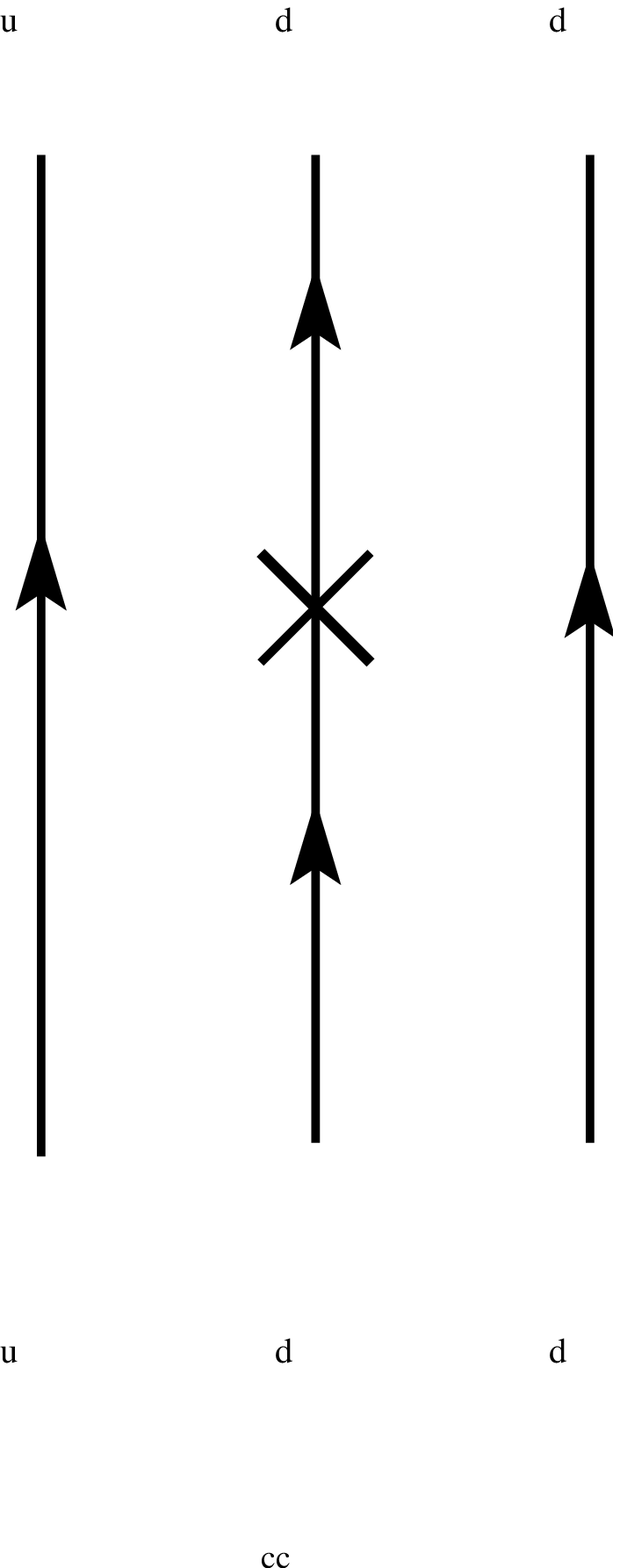}
\hspace{3.3cm}
\includegraphics[height=60mm]{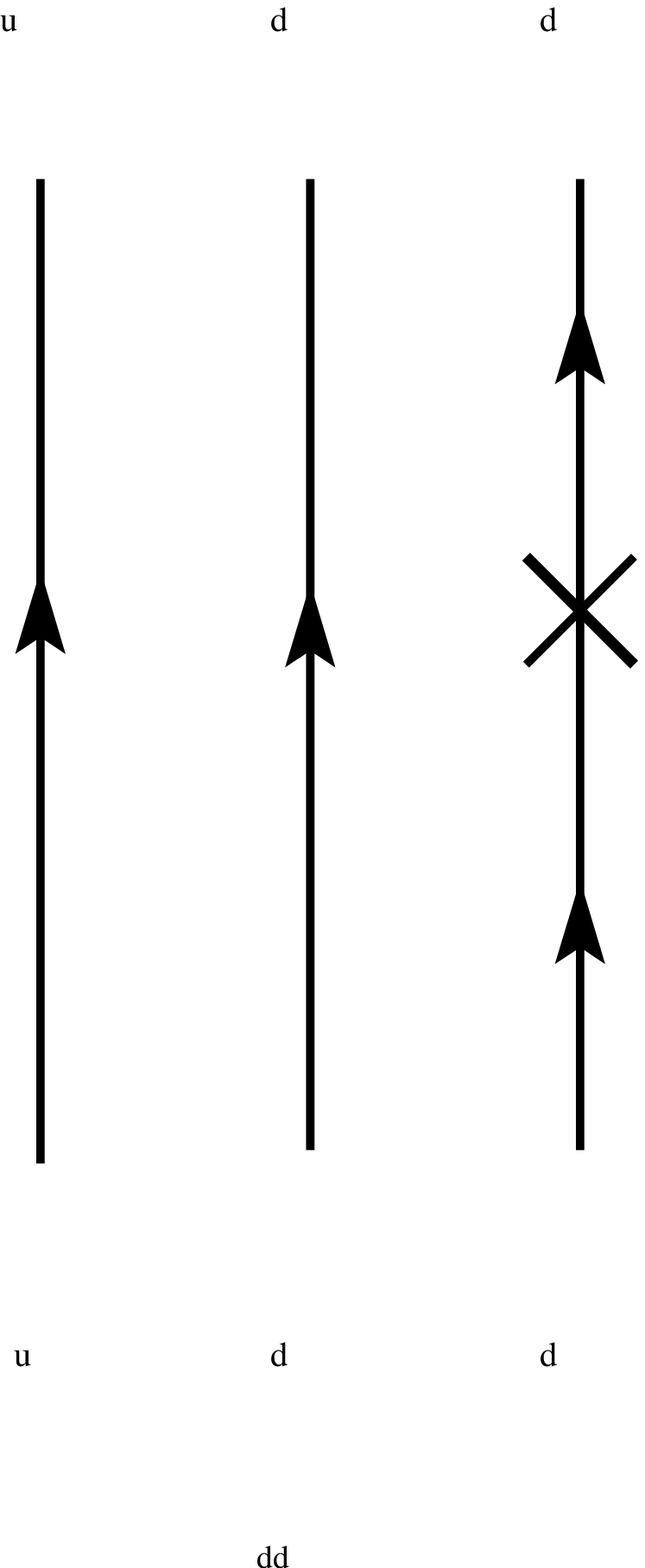}\\
\vspace{1.2cm}
\hspace{-.1cm}
\includegraphics[height=60mm]{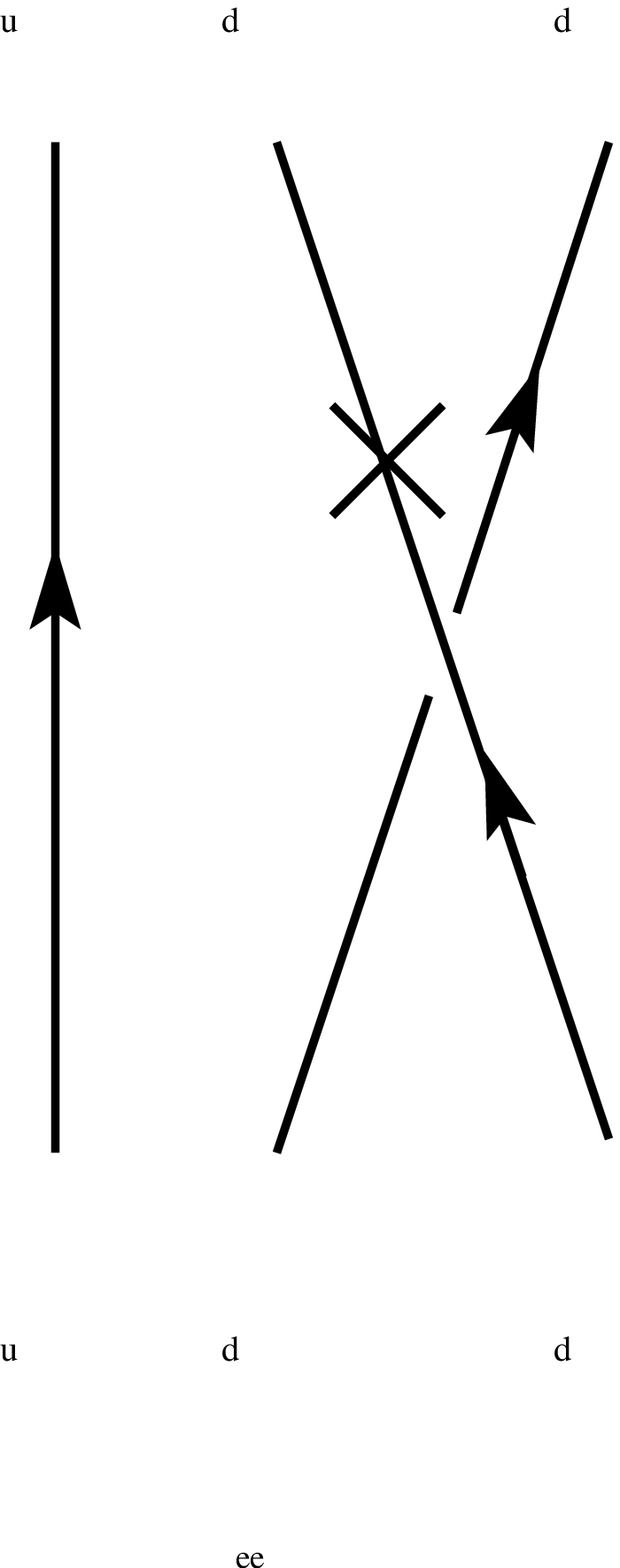}
\hspace{3.3cm}
\includegraphics[height=60mm]{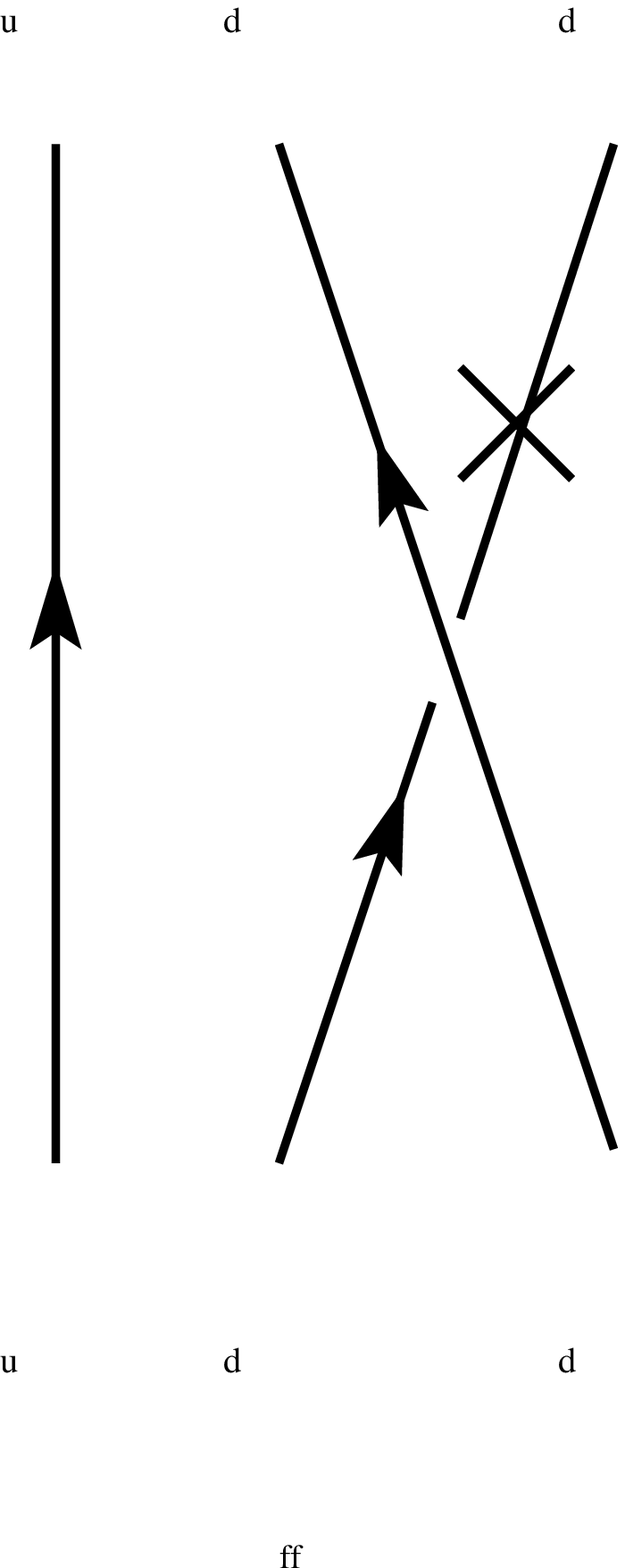}
\Ec
\caption{Six diagrams contributed to the neutron three-point function.  
Cross symbol represents the current insertion. In $(a),\,(b)$ EM
current  
is connected with $u$ quark line, and in $(c)\sim (f)$ with $d$ quark
line.}
\label{diag_u}
\end{figure}

To compute nucleon three-point functions, we construct the amplitudes
for the six diagrams 
in Fig.\ref{diag_u},  each of which consists of two
quark propagators and one double quark propagator with an insertion of
the 
electromagnetic current $J_{\mu}^{\rm EM}$:
\BE
  J_{\mu}^{\rm EM}(x) = e^u V^u_{\mu}(x) + e^d V_{\mu}^d(x) + e^s
V_{\mu}^s(x)
\EE
with $V_{\mu}^i(x)\,(i=u,d,s)$ the conserved vector current of flavor
$i$ defined by 
\BA
V^i_{\mu}(x) &=& \sum_s  j^i_{\mu}(x,s), \\
  j^i_{\mu}(x,s) &=&
\frac{1}{2}\Big[\bar\psi^i(x,s)(1-\gm_{\mu})U_{\mu}(x)\psi^i(x+\hat\mu,s)\nn
            &&+
\bar\psi^i(x+\hat\mu,s)(1+\gm_{\mu})U_{\mu}^{\dag}(x)\psi^i(x,s)\Big].
\EA
Since this current is conserved,  the 
renormalization factor satisfies $Z_V=1$.
To insert $V_{\mu}^i$, we use the ordinary source method. 

\begin{figure}[h]
\Bc
\psfrag{t2}{ }
\psfrag{tau}{ }
\psfrag{t1}{ }
\psfrag{u}{\Large u}
\psfrag{d}{\Large d}
\psfrag{ud}{\Large u, d, s}
\includegraphics[width=80mm]{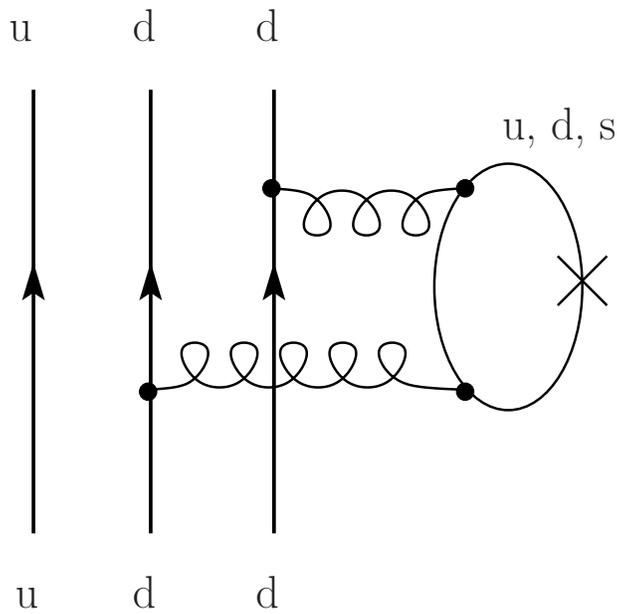}
\Ec
\caption{An example for the disconnected contribution to  
three-point function. 
The loop contains $u$, $d$, $s$ quarks.}
\label{discdiag}
\end{figure}

We do not consider the 
disconnected diagrams depicted in Fig.~\ref{discdiag},
since this contribution vanishes in the SU(3) flavor symmetric limit due
to the fact that
$e^u+e^d+e^s=0$. 
In future investigations, however, it is necessary to check
if this contribution is indeed small or not for $m_{u,d}\not= m_s$.


\subsection{Lattice parameters}
We generate quenched gauge configurations on a $16^3\times 32$ lattice
with 
the RG improved gauge action at $\bt=2.6$.
Gauge configurations separated by 200 sweeps are used for measurements
after 2000 sweeps for thermalization, where one sweep consists of one 
heat-bath update followed by four over-relaxation steps.
The quark propagator is calculated using the domain-wall fermion with
$N_s=16$
and $M=1.8$, 
where $N_s$ is a size of the fifth dimension and $M$ is the domain-wall 
height. 
These parameters are the same as those employed 
in Ref.\cite{BK_CP-PACS},
in which the lattice spacing was determined to be $a^{-1}=1.875(56)$GeV
from
the $\rho$ meson mass.   
The residual quark mass $m_{\rm res}$, which represents the explicit
chiral 
symmetry breaking effect at finite $N_s$, 
is found to be $m_{\rm res}\simeq 4$MeV.

We take the bare quark mass $m_fa = 0.03$ corresponding to the ratio of 
pseudoscalar to vector meson masses 
$m_{PS}/m_{V}=0.629(8)$\cite{BK_CP-PACS}. 
To reduce the effects of unwanted excited hadron states,
we calculate the quark propagator employing an exponentially smeared
source,
given by $Ae^{-Bra}$ with 
$A=1.28$ and $B=0.40$, where
$ra$ is the distance from the origin at $(7,7,7)$ in the lattice unit.
We take the periodic boundary condition for both gauge and fermion
fields 
in order to minimize possible violation of the equivalence between 
the bosonic and the fermionic definitions 
of topological charge.

In the calculation of three-point functions, the electromagnetic current
is 
inserted at the fixed time slice $\tau=6$.
The nucleon sink is varied as a function of $t$ with
the nucleon source fixed at $t=1$.
We move the final state nucleon with one of the three non-zero spatial
momenta
${\vec p}=(\pi/8,0,0),\,(0,\pi/8,0),\,(0,0,\pi/8)$, 
while the initial state nucleon is placed at rest $\vec{p^\prime}
=(0,0,0)$. 

We accumulate 730 configurations for measurements. 
Errors are estimated by the single elimination jackknife procedure 
for all measured quantities.

\section{Numerical results \label{sec:result}}

\subsection{Topological charge}
\begin{figure}[h]
\Bc
\psfrag{#conf}{\large \hspace{-1cm}\#cooling step}
\includegraphics[width=100mm, angle=0] {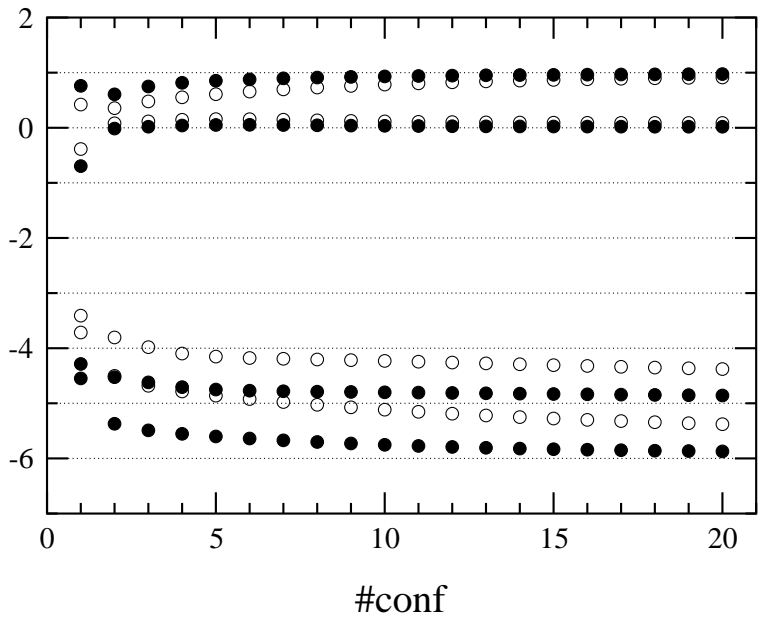}
\Ec
\caption{Topological charge as a function 
of the number of cooling steps. Open symbols represent the naive
plaquette 
definition and solid symbols the improved  definition.}
\label{cool20}
\end{figure}

In order to reduce ultra-violet fluctuations of gauge configurations in
the calculation of the
topological charge, we employ the cooling method using the action  in
eq.(\ref{eq:Gauge_action})
 with $c_1= 3.648$ and $c_0=-0.331$.
In Fig.~\ref{cool20} we show both $Q_{\rm plaq}$ and $Q_{\rm imp}$
as a function of the cooling steps,
where $Q_{\rm plaq}$ is the naive plaquette definition 
of the topological charge and $Q_{\rm imp}$ is the improved one
in eq.(\ref{top_def}).
It is clearly observed that $Q_{\rm imp}$
approach integer values much faster than $Q_{\rm plaq}$ as the cooling
step 
increases.  The values after 20 cooling steps are also closer 
to integer values.
The deviation from integer is within a few percent for 
$Q_{\rm imp}$ after 20 cooling steps. As the topological charge,
we employ 
$Q_{\rm imp}$ measured after 20 cooling steps on each configuration.

\begin{figure}[h]
\Bc
\vskip 10mm 
\includegraphics[width=120mm, angle=0] {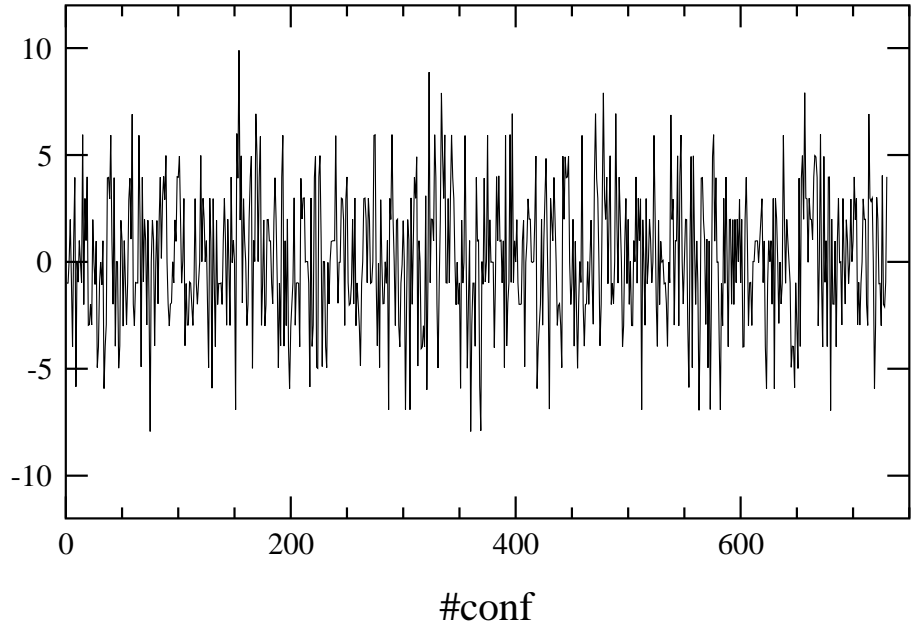}
\Ec
\caption{Time history of the topological charge.
Gauge configurations are separated by 200 sweeps.}
\label{conf_topo}
\end{figure}

\begin{figure}[h]
\Bc
\vskip 10mm 
\includegraphics[width=120mm, angle=0] {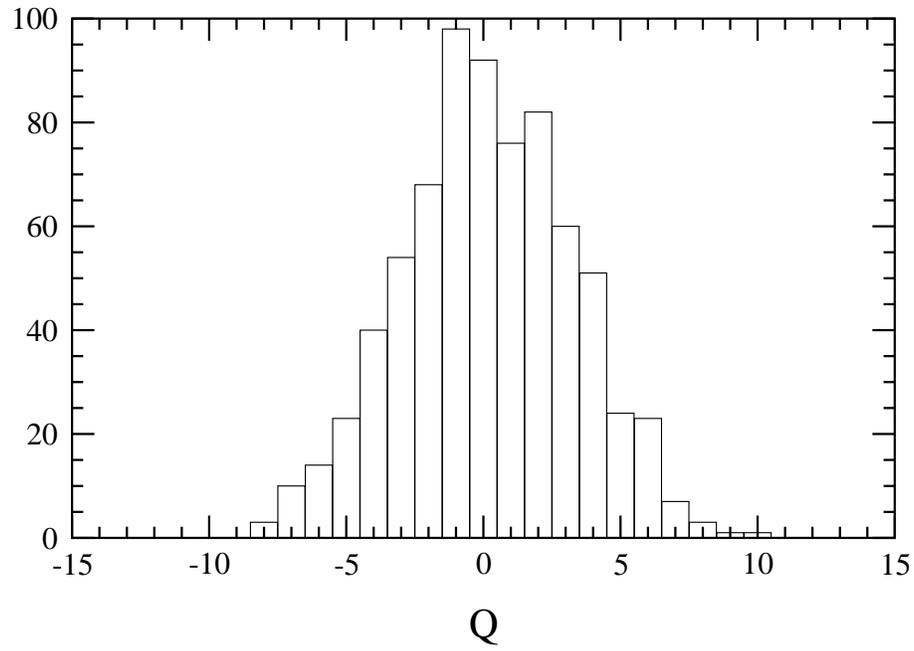}
\Ec
\caption{Histogram of the topological charge.}
\label{hist}
\end{figure}

In Fig.~\ref{conf_topo} we show the time history of the topological
charge as a function of the gauge configuration separated by 200 sweeps,
where we do not recognize any long autocorrelation.
In addition, the histogram of the topological charge 
in Fig.~\ref{hist} exhibits nearly a Gaussian distribution.
These observations suggest that the sampling of the topological charge
is 
sufficiently good in our calculation.


\subsection{Nucleon propagator}
\subsubsection{$O(\theta^0)$ contribution}
{}From the analysis in Sec.~\ref{sec:formula_2pt}, 
the $O(\theta^0)$ contribution of the
nucleon propagator should behave as
\BE
  \tr[G_{NN}({\vec p},t)\Gamma_4] = \LL\{ 
\begin{array}{cc}
\vert Z_N\vert ^2 e^{-m_Nt}+\cdots & (|{\vec p}|=0) \\
\vert Z_N(p)\vert^2\frac{m_N}{E_N}e^{-E_Nt} + \cdots & (|{\vec p}|=1)
\end{array}
\RR. ,
\EE
where $
  \Gamma_4 = (1+\gm_4)/2 $ 
and the dots represent the contribution from the excited states.
Here $\vert{\vec p}\vert =1$ is a shorthand notation of
${\vec p} = (\pi/8) {\vec n}$ with $\vert{\vec n}\vert = 1$.

\begin{figure}[h]
\Bc
\includegraphics[width=100mm, angle=0] {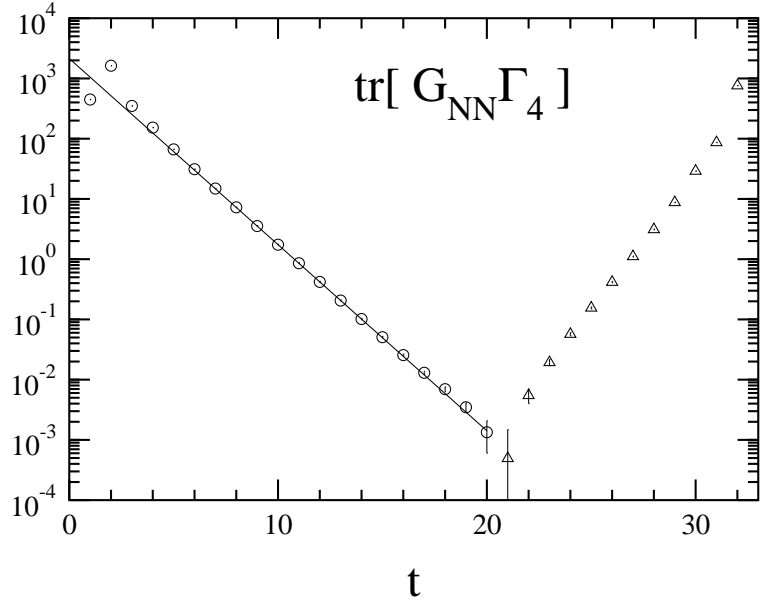}
\vskip 10mm 
\includegraphics[width=100mm, angle=0] {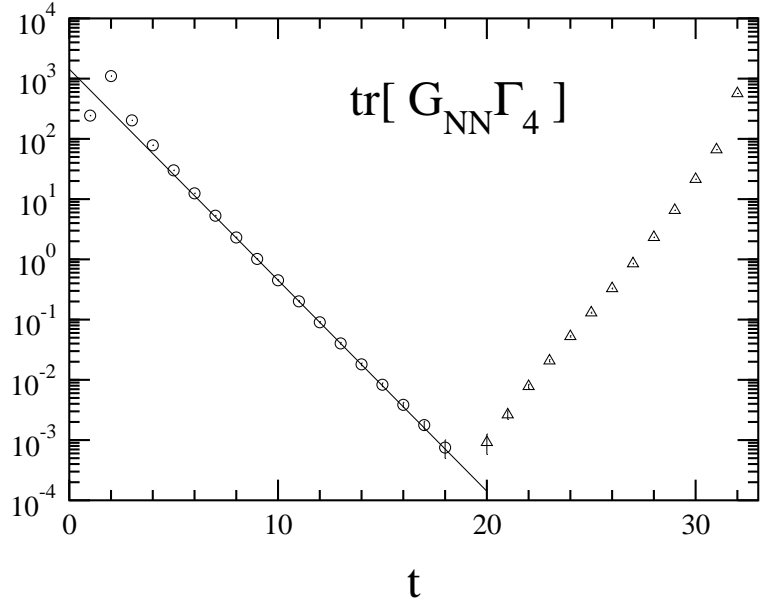}
\Ec
\caption{$\tr\left[ G_{NN}(\vec{p},t)\Gamma_4\right]$,
$O(\theta^0)$ contribution of the nucleon propagator 
with $\Gamma_4$ projection for $|{\vec p}|=0$ (top) 
and $|{\vec p}|=1$ (bottom).
Solid lines denote the fitting results. 
The parity-odd state dominates the backward propagation in $t$ 
denoted by triangles.
}
\label{Npp}
\end{figure}

\begin{figure}[h]
\Bc
\includegraphics[width=100mm, angle=0] {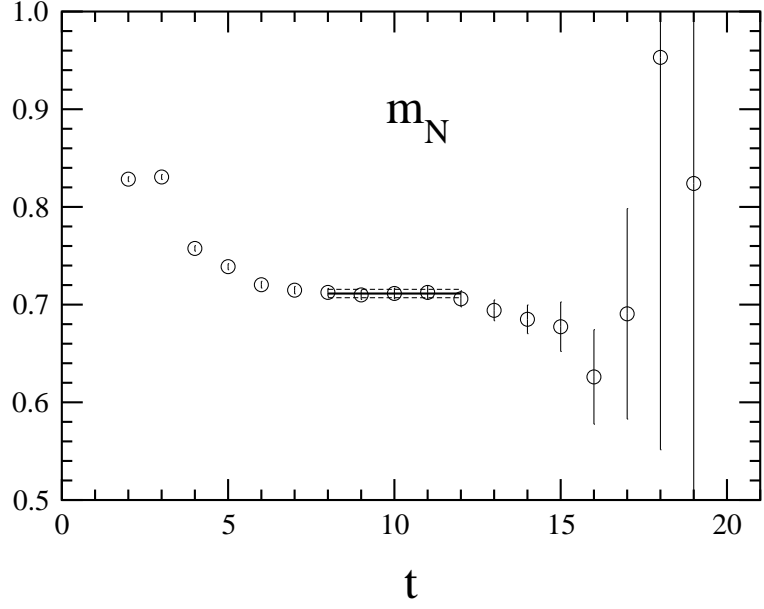}
\vskip 10mm 
\includegraphics[width=100mm, angle=0] {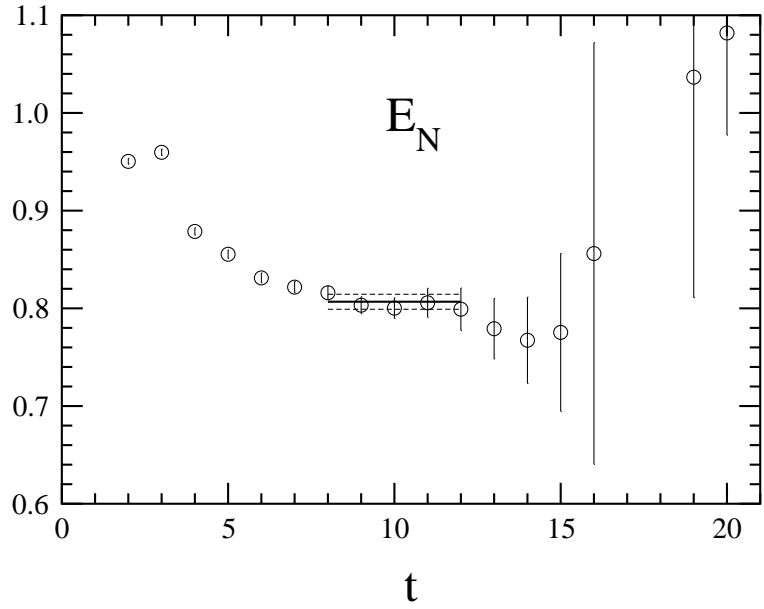}
\Ec
\caption{Effective mass plot for the nucleon with $|{\vec p}|=0$ (top)
and $|{\vec p}|=1$ (bottom). 
Solid(dotted) lines denote the central values(errors) of the global
fit.}
\label{Nem}
\end{figure}

The nucleon propagators from the smeared source and the point sink 
with $|{\vec p}|=0$ and $|{\vec p}|=1$ are plotted in Fig.~\ref{Npp},
and the corresponding effective masses are given in Fig.\ref{Nem}. 
Choosing $9\le t\le 13$ for the fitting range,
we obtain
\BA
  E_Na &=& \LL\{
\begin{array}{cc}
 0.7114(43) & (|{\vec p}|=0) \\
 0.8068(77) & (|{\vec p}|=1)
\end{array}
\RR. ,
\NN
 Z_N &=& \LL\{
\begin{array}{cc}
 2162(87) & (|{\vec p}|=0) \\
 1453(104) & (|{\vec p}|=1)
\end{array}
\RR.  .
\EA
Note that $E_N a = 0.8068(77)$ at $\vert\vec{p}\vert =1$ agrees with
the relativistic dispersion relation
$\sqrt{(m_Na)^2+(\pi/8)^2}=0.813(4)$
within the error.


\subsubsection{$O(\theta^1)$ contribution and the determination of
$f_N^1$}
\begin{figure}[h]
\Bc
\includegraphics[width=100mm, angle=0] {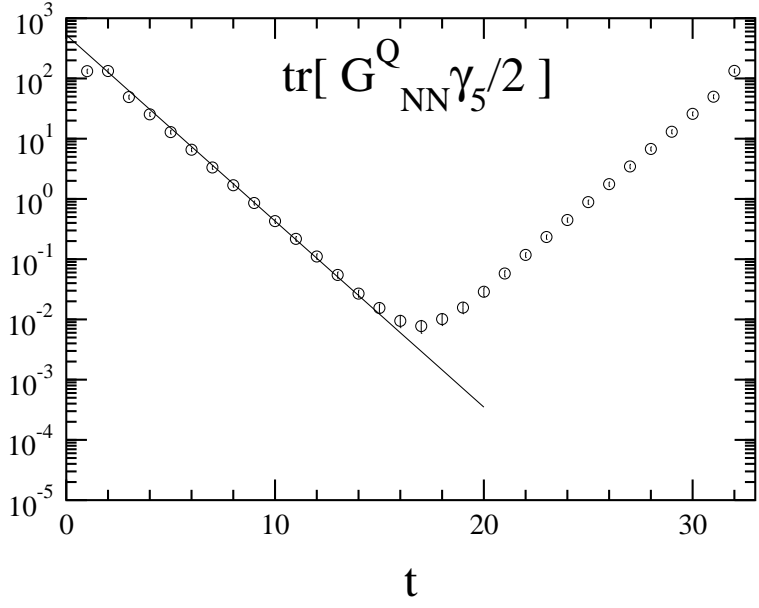}
\vskip 10mm 
\includegraphics[width=100mm, angle=0] {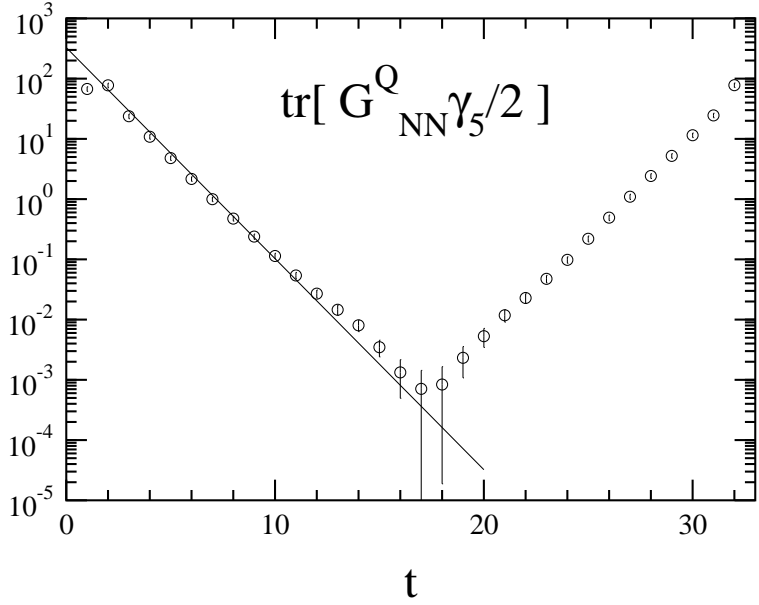}
\Ec
\caption{$\tr\left[ G_{NN}^Q(\vec{p},t)\frac{\gamma_5}{2} \right]$,
parity odd part of the nucleon propagator with $Q$ for
eq.(\ref{eq:odd_part0}) 
$|{\vec p}|=0$ (top) and eq.(\ref{eq:odd_part}) $|{\vec p}|=1$ (bottom).
The sign of 
   propagator is negative. Solid lines denote the fitting results.}
\label{Npp_Q_gm5}
\end{figure}

\begin{figure}[h]
\Bc
\includegraphics[width=100mm, angle=0] {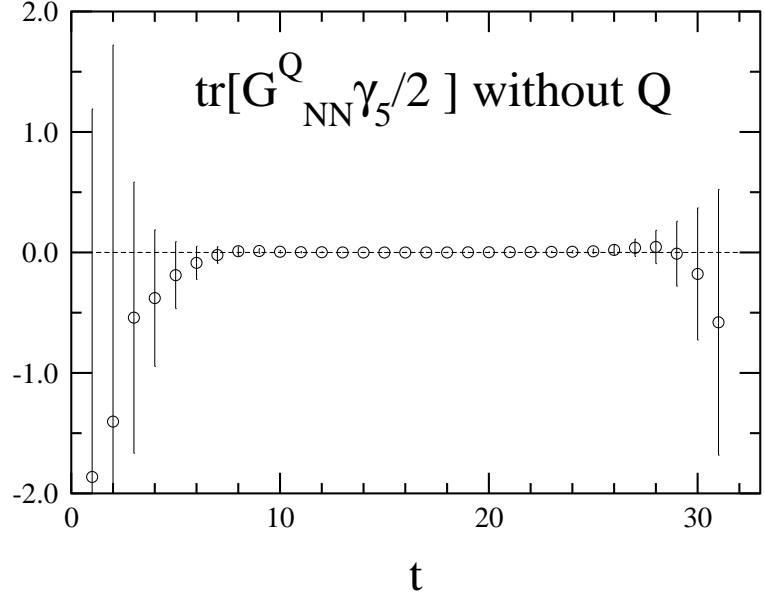}
\vskip 10mm 
\includegraphics[width=100mm, angle=0] {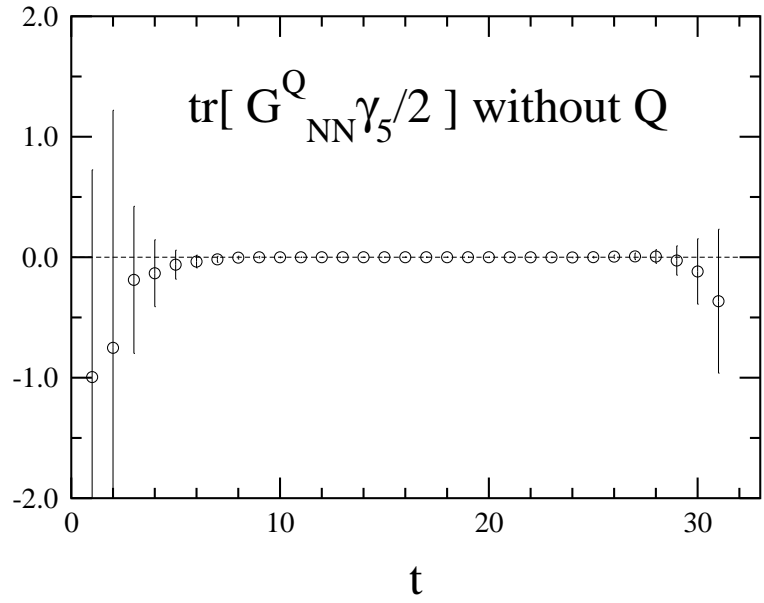}
\Ec
\caption{$\tr\left[ G_{NN}(\vec{p},t)\frac{\gamma_5}{2} \right]$,
parity odd part of the nucleon propagator without $Q$
for eq.(\ref{eq:odd_part0}) $|{\vec p}|=0$ (top) and
eq.(\ref{eq:odd_part}) 
$|{\vec p}|=1$ (bottom).}
\label{Npp_noQ_gm5}
\end{figure}

In Sec.~\ref{sec:formula_2pt} we have shown that 
the parity-odd part of the nucleon propagator should have the following
form
at the order of $\theta^1$: 
\BA
  \tr[G_{NN}^Q(0,t)\frac{\gm_5}{2}] &=& |Z_N|^2f_N^1
e^{-m_Nt},\label{eq:odd_part0}\\
  \tr[G_{NN}^Q(p,t)\frac{\gm_5}{2}] &=& |Z_N(p)|^2\frac{m_N}{E_N}f_N^1
e^{-E_Nt}.
\label{eq:odd_part}
\EA
In Fig.\ref{Npp_Q_gm5} we plot the parity-odd part with the insertion of
$Q$,
while the parity-even part without $Q$,
$\tr[ G_{NN}(p,t) \frac{\gm_5}{2}]$, is given in Fig.\ref{Npp_noQ_gm5}
for comparison.
The latter vanishes at all time slices as expected.
On the other hand, the former is non-zero,
showing an exponential fall-off at large $t$.
After averaging over the forward and backward propagators in time,
we fit the party-odd part at $9\le t\le 14$, by the form
$\xi e^{-E_N t}$ with $E_N$ fixed to the value obtained from the
previous fit 
of the $O(\theta^0)$ contribution. 
The fit, represented by the solid line in Fig.~\ref{Npp_Q_gm5},
is reasonably good, showing that
the decay rate of the parity-odd part is consistent with that of the
parity-even part for both $|{\vec p}|=0$ and 1.
Note however that the small deviation indicates that the present 
statistics are still insufficient at large $t$ for $|{\vec p}|=1$.

\begin{figure}[h]
\Bc
\includegraphics[width=100mm, angle=0] {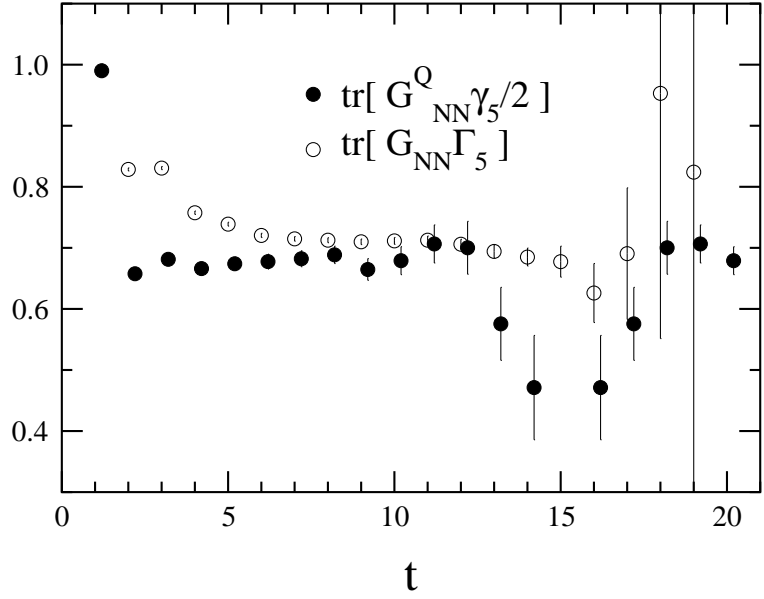}
\vskip 10mm 
\includegraphics[width=100mm, angle=0] {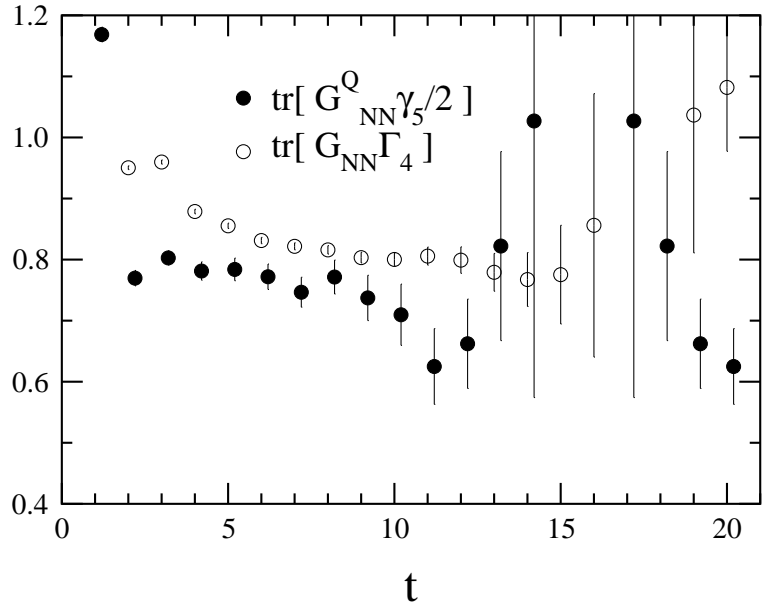}
\Ec
\caption{Effective mass plot for the parity odd part of the nucleon
propagator with $Q$ (filled) for eq.eq.(\ref{eq:odd_part0}) $|{\vec
p}|=0$ (top) 
and eq.(\ref{eq:odd_part}) $|{\vec p}|=1$ (bottom) after averaged with
backward propagator 
in time direction. 
The parity even part without $Q$ (open), which are already shown
in Fig.~\ref{Nem}, are also plotted for comparison.}
\label{Neff_Q_gm5}
\end{figure}

Another confirmation of consistency is found in the effective mass plot 
of Fig.~\ref{Neff_Q_gm5}, where the parity-odd part of the nucleon
propagator with $Q$ and the parity-even part without $Q$ are compared.

With the aid of $E_N$ and $Z_N$ obtained previously, $\xi$ 
is converted to $f_N^1$ as
\BE
  f^1_N = \LL\{
\begin{array}{cc}
 -0.247(17) & (|{\vec p}|=0) \\
 -0.243(20) & (|{\vec p}|=1)
\end{array} \RR. .
\EE
Despite the deviation at large $t$ for $|{\vec p}|=1$, 
the two estimates of $f_N^1$ agree within the errors.
In our analysis we will use $f_N^1$ obtained from $|{\vec p}|=0$.

We note that the exponential fall-off of $G_{NN}^Q$ 
is very sensitive to the
distribution of the topological charge. 
One could use this quantity to judge
whether the sampling of the topological charge is sufficiently good:
if the exponential fall-off of $G_{NN}^Q$ does not agree with that of
$G_{NN}$,
the sampling is not satisfactory.


\subsection{Form factors}
\subsubsection{Extraction from three-point functions}

\begin{figure}[h]
\Bc
\includegraphics[width=100mm, angle=0] {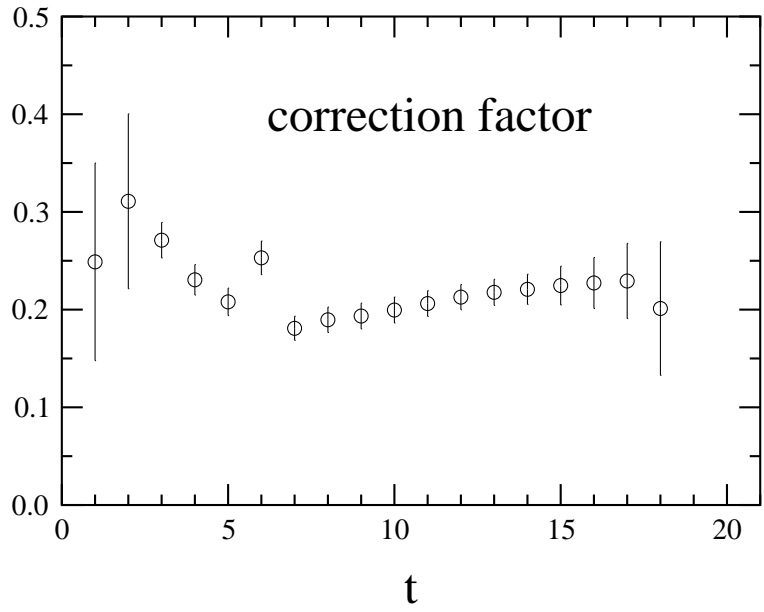}
\Ec
\caption{Correction factor $R(q;\tau,t)$ as a function of $t$ with
$\tau=6$ and
$\vert {\vec q}\vert=1$.}
\label{corr}
\end{figure}

In order to remove the exponential fall-off due to the
nucleon propagation in the three-point functions,
we divide them by the two-point function
$G_{NN}({\vec p},t)$ with the smeared source at $t=1$ 
and the point sink at $t$:
\BA
 \Pi^e_{\mu}(q;t,\tau) &=&
\frac{G_{NJ_{\mu}N}(q;t,\tau)}{\tr[\Gamma_4 G_{NN}(q,t)]}R(q;\tau,t), 
 \label{3pt_ov_nnE}\\
 \Pi^o_{\mu}(q;t,\tau) &=& \frac{G_{NJ_{\mu}N}^Q(q;t,\tau)}{\tr[\Gamma_4 G_{NN}(q,t)]}
R(q;\tau,t)\label{3pt_ov_nnO}
 \EA
with
\BE
R(q;\tau,t) = \LL[\frac{\tr[\Gamma_4 G_{NN}(q,t)]\tr[\Gamma_4G_{NN}(q,\tau)]
\tr[\Gamma_4 G^{PP}_{NN}(0,t-\tau)]}
{\tr[\Gamma_4 G_{NN}(0,t)]\tr[\Gamma_4 G_{NN}(0,\tau)]\tr[\Gamma_4G^{PP}_{NN}(q,t-\tau)]}\RR]^{1/2},
\EE
where the factor $R$ is to remove the interpolating field 
dependent normalization factors.
Here we introduce a new two-point function
$G^{PP}_{NN}(q,t)$ that has the point source at 
$t=1$ and the point sink at $t$,
and $G^{PP}_{NN}(q,t-\tau)$ represents that the source point
is started at $t=\tau$. 
In Fig.~\ref{corr} we plot the $t$ dependence of $R(q;\tau,t)$ 
with $\tau=6$ and $\vert {\vec q}\vert=1$.
As we expect,
the correction factor $R$ is independent of $t$ in the large $t$ region.


The form factors are extracted from the ratios defined in
eqs.(\ref{3pt_ov_nnE}) and (\ref{3pt_ov_nnO}) by applying appropriate
projections of gamma matrices. From the $O(\theta^0)$ 
contribution $\Pi^e_4(q;t,\tau)$ of eq.(\ref{3pt_ov_nnE}),
$F_1$ and $F_2$ can be extracted as
\BA
  \tr[\Pi^e_4(q;t,\tau)\Gamma_4] &=& \frac{E_N+m_N}{E_N}
  \LL(F_1(q^2) + F_2(q^2)\frac{q^2}{4m^2_N}\RR), \label{elec_mom}\\
  \tr[\Pi^e_i(q;t,\tau)i\Gamma_4\gm_5\gm_j] &=&
\eps_{ijk}\frac{q_k}{E_N}\LL(F_1(q^2) + F_2(q^2)\RR),
  \label{magne_mom}
\EA
where we use eq.(\ref{eq:3pt_even}).
The $F_1$ and $F_2$ are related to the electric ($G_e$) 
and the magnetic ($G_m$) form factors as
\BA
  G_e(q^2) &=& F_1(q^2) + F_2(q^2)\frac{q^2}{4m^2_N} \label{Ge_f12},\\
  G_m(q^2) &=& F_1(q^2) + F_2(q^2)\label{Gm_f12} .
\EA
These quantities become the electric charge and the magnetic moment 
at $q^2\rightarrow 0$. Therefore
$G_e(0)=0(1)$ and $G_m(0)=-1.91(2.79)$\cite{pdg2004}
for the neutron(proton). 

We can extract the form factor $F_3(q^2)$, which is relevant 
to the electric dipole moment,
in two different ways  from  $\Pi^o_4(q;t,\tau)$ of
eq.(\ref{3pt_ov_nnO}):
\BA
\tr[\Pi_4^o(q,t,\tau)\Gamma_{4}\gm_5]
&=& \frac{{\vec q}^2}{2E_Nm_N}F_3(q^2) \nn
&+& \bigg[\frac{E_N+m_{N}}{2E_N}F_1(q^2) 
   + \frac{{\vec q}^2}{4m_NE_N}F_2(q^2)\bigg]f^1_N, \label{f3_proj1}\\ 
\tr[\Pi_4^o(q,t,\tau)i\Gamma_{4}\gm_5\gm_i] &=&
-\frac{E_N+m_N}{2E_Nm_N}q_iF_3(q^2)\nn
&+&  \LL[-\frac{q_i}{2E_N}F_1(q^2)
    -\frac{q_i(E_N+3m_N)}{4m_NE_N}F_2(q^2)\RR]f^1_N\label{f3_proj2} ,
\EA
where we use eq.(\ref{eq:3pt_odd}). 
It will be clear that the consistency between the two different 
extractions of $F_3(q^2)$ provides a crucial test for the validity of
actual calculations. Note that the anapole form factors which appears in
$W_{\mu}^{\rm odd}(q)$ 
is canceled out after applying the trace of spinor with these projections.

\subsubsection{Parity-even part}

\begin{figure}[h]
\Bc
\vskip 10cm 
\includegraphics[width=100mm, angle=0] {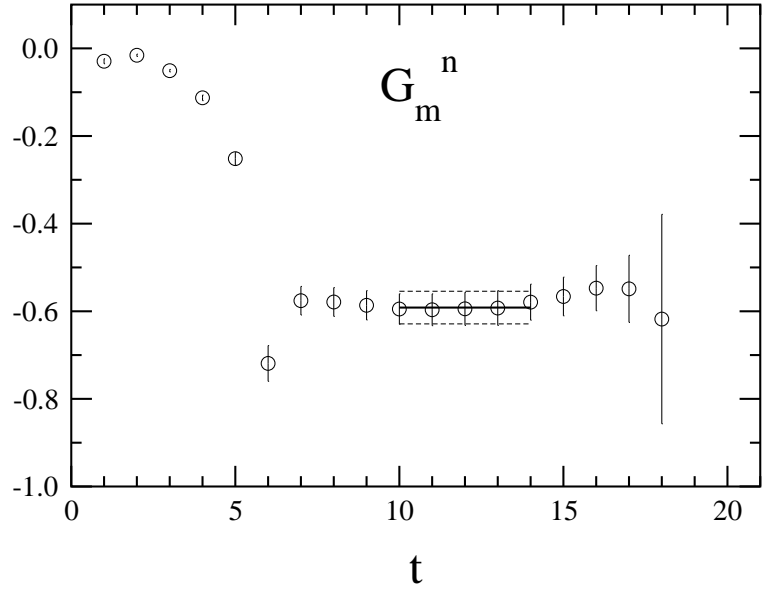}
\vskip 10mm 
\includegraphics[width=100mm, angle=0] {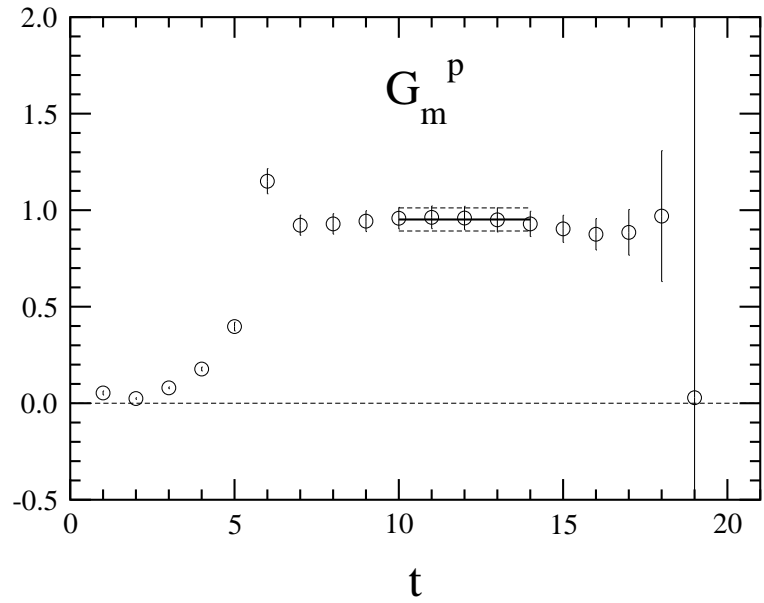}
\Ec
\caption{Magnetic form factor for neutron (top) and proton (bottom). }
\label{Gm}
\end{figure}

\begin{figure}[h]
\Bc
\includegraphics[width=100mm, angle=0] {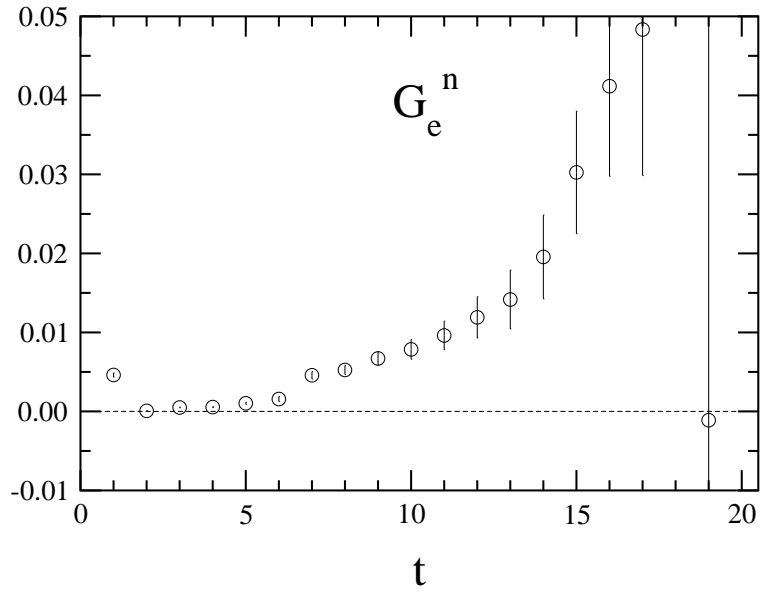}
\vskip 10mm 
\includegraphics[width=100mm, angle=0] {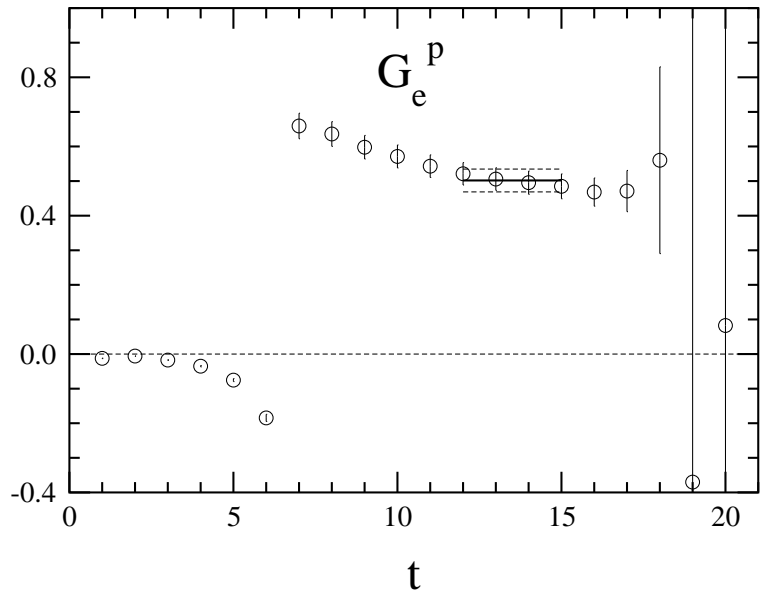}
\Ec
\caption{Electric form factor for neutron (top) and proton (bottom). }
\label{Ge}
\end{figure}

We first consider the parity-even part of the form factor.
In Figs.~\ref{Gm} and \ref{Ge} we show the 
magnetic and electric form factors obtained from 
eqs.(\ref{elec_mom}) and (\ref{magne_mom}).
We observe good plateaux 
except for the electric form factor of the neutron.
Employing a constant fit with $10\le t\le 14$ for $G_m$ and $12\le t\le
15$ for $G_e$ 
we obtain
\BA
  G_m(q^2) &=& \LL\{
\begin{array}{cl}
  -0.591(37) & \textrm{(neutron)}\\
   0.952(60) & \textrm{(proton)}
\end{array}
\RR. \\
G_e(q^2) &=& 
   0.502(33) \quad \textrm{(proton)}
\EA
at $q^2 = 0.58$ GeV$^2$. Here the value for $G_e^n$ is not quoted since 
it is difficult to find a plateau for the fit. Note also that the value
of $G_e$ slightly depends on the fitting range, 
as could be seen from the figure.
These values are compared with the previous lattice result\cite{QCDSF}
obtained at a similar nucleon mass and lattice spacing 
using the nonperturbatively $O(a)$ improved 
Wilson quark and the plaquette gauge actions:
$ G_m(q^2) =  -0.73(4)$  for the neutron, and
$ G_m(q^2) =   1.17(6)$ and $G_e(q^2) =  0.451(13)$ for the proton.

Experimentally the proton electromagnetic form factors at small $q^2$
have 
been measured as\cite{Price},
\BE
  G_m^{p({\rm exp})}(q^2=0.58{\rm GeV}^2) = 0.848\pm0.011,\quad 
  G_e^{p({\rm exp})}(q^2=0.58{\rm GeV}^2) = 0.294\pm0.008.
\EE
Recent experimental measurements for neutron form factors cover a wide 
range of 
$q^2$ from very small $q^2\sim 0.3$ GeV$^2$ to 
more than $1$ GeV$^2$ with high precision\cite{Gao}. 
For example, the neutron magnetic form factor at $q^2=0.6$ GeV$^2$ is 
measured as\cite{Xu} 
\BE
  G_m^{n(\rm{exp})}(q^2=0.6\rm{ GeV}^2) = -0.568 \pm 0.007 \rm{
(stat.)}\pm0.015\rm{ (syst.)},
\EE
and the electric form factor at $q^2=0.5$ GeV$^2$ is\cite{Zhu}  
\BE
  G_e^{n(\rm{exp})}(q^2=0.5\rm{ GeV}^2) = 0.0463\pm0.0062\rm{
(stat.)}\pm0.0034\rm{ (syst.)}.
\EE
The fact that our results are not so much different from these
experimental 
values may suggest that
systematic errors involved in our calculations, such as 
the large quark mass, the small lattice size, the non-zero lattice
spacing 
and the quenching, are not so large on these quantities.

\begin{figure}[h]
\Bc
\includegraphics[width=100mm, angle=0] {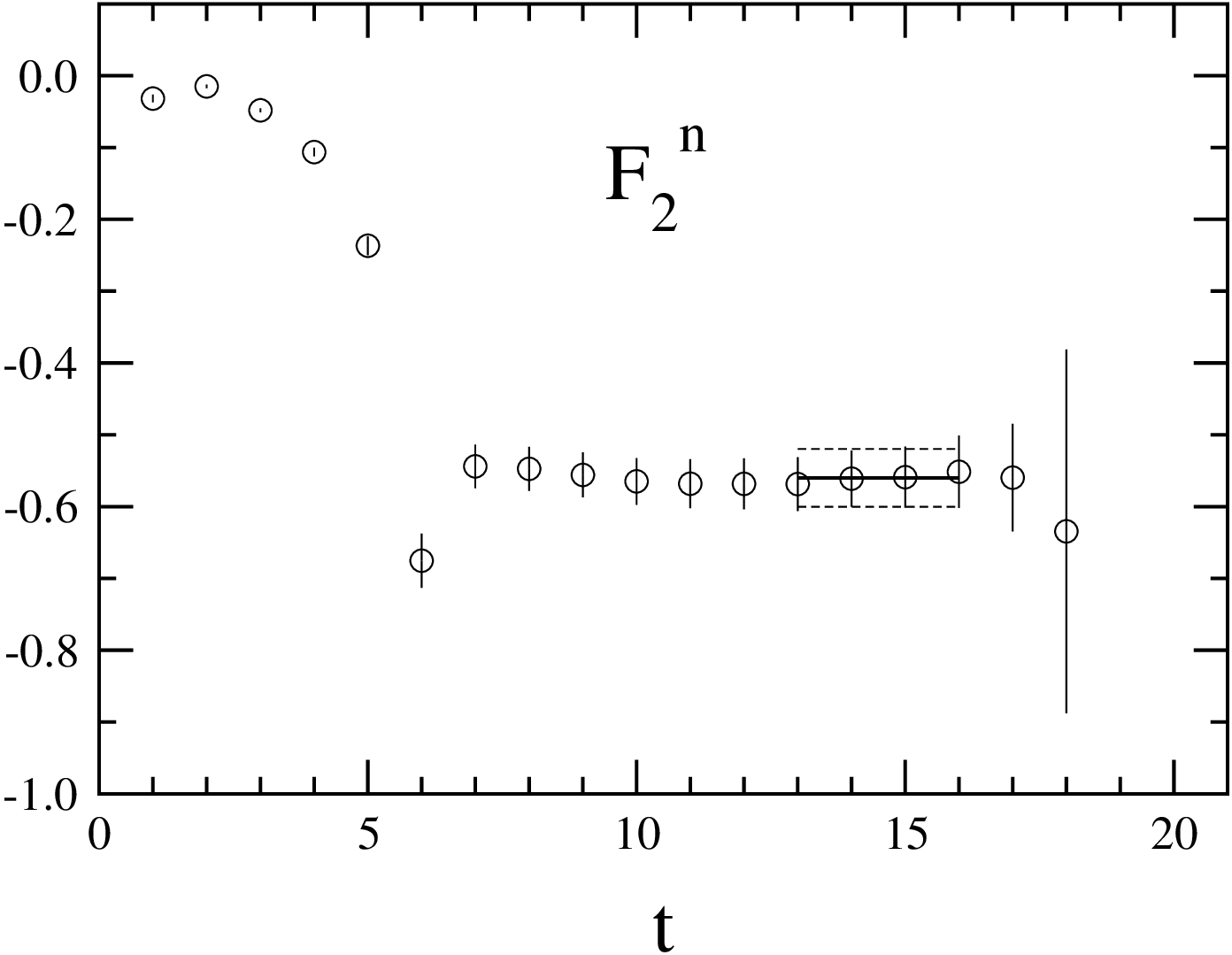}
\vskip 10mm 
\includegraphics[width=100mm, angle=0] {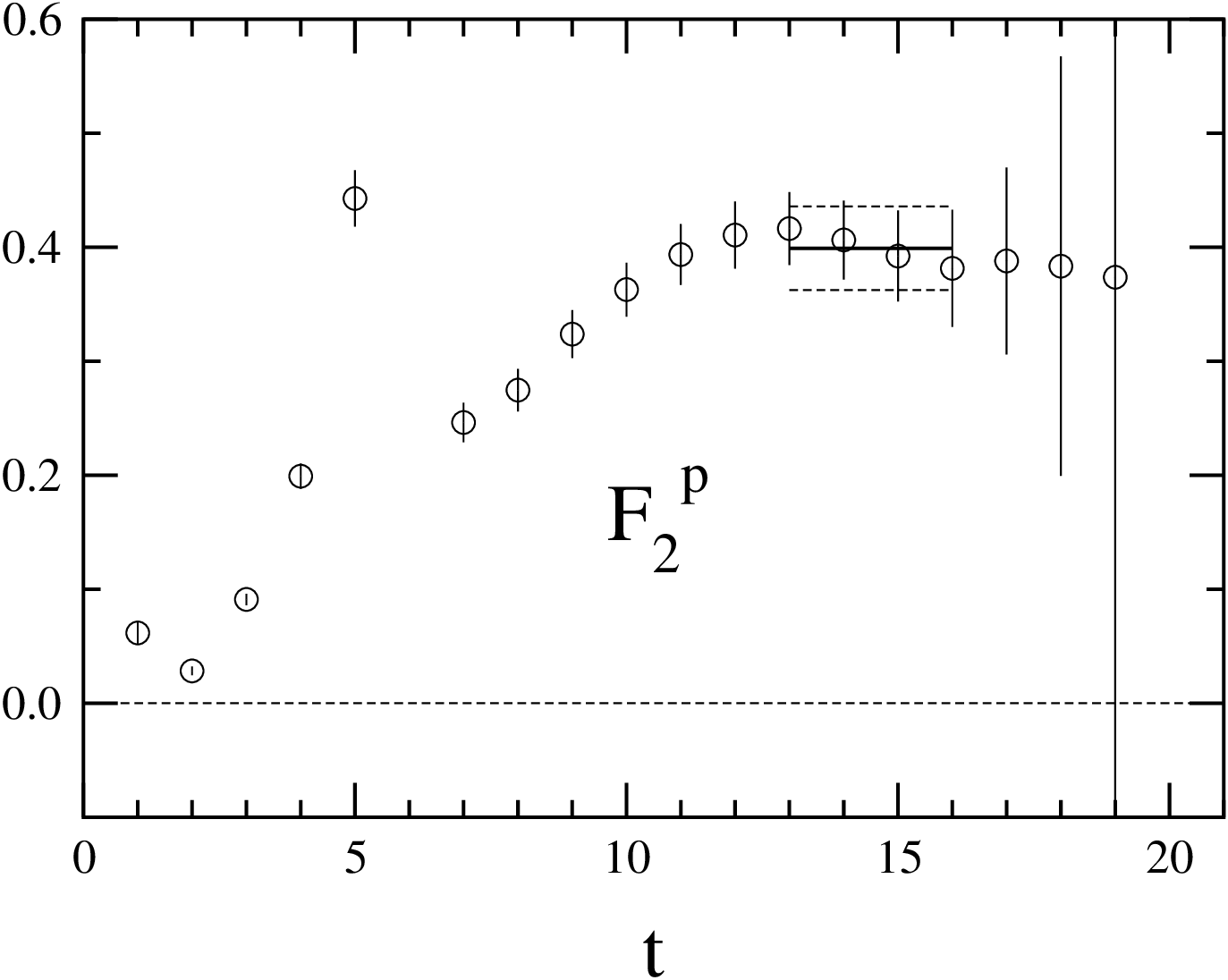}
\Ec
\caption{Form factor  $F_2(q^2)$ for neutron (top) and proton (bottom).
Horizontal lines represent the constant fit.}
\label{f2}
\end{figure}

The form factors $F_1$ and $F_2$ are obtained from $G_m$ and $G_e$
using eqs.(\ref{Ge_f12}) and (\ref{Gm_f12}). 
We show the time dependence of $F_1$ and $F_2$
in Figs.~\ref{f1} and \ref{f2}.
We observe plateaus at $t \ge 13$ for $F_2^{n,p}$ and $F_1^p$ but not
for 
$F_1^n$, which behaves badly as $G_e^n$ does.
We think that this behavior is caused by  
the contribution from other excited states, as pointed out 
in Ref.\cite{Tang}. 
Therefore we have tried to include the contributions of 
the parity-odd nucleon state $N_-$ in our analysis,
using the formulae given in Appendix~\ref{app1}.
Although larger errors prevent us from extracting the reliable values
for $G_e^n$ and $F_1^n$,
we find a much better plateau for $G_e^n$ as shown in Fig.~\ref{ge_nodd}
and for $F_1^n$ in Fig.~\ref{f12_nodd}.

Applying a constant fit to $F_1^{p}$ and $F_2^{n,p}$ with $13\le t\le
16$,
we obtain
 \BE
 F_1^p(q^2) = 0.515(37),\,\,  F_2^n(q^2) = -0.560(40),\,\,  
F_2^p(q^2) = 0.399(37), 
 \EE
which are compared with 
$F_1^p(q^2)=0.499(13)$, $F_2^p(q^2)=0.68(6)$
from Ref.\cite{QCDSF}. 


\subsubsection{Parity-odd part}

\begin{figure}[h]
\Bc
\includegraphics[width=100mm, angle=0]
{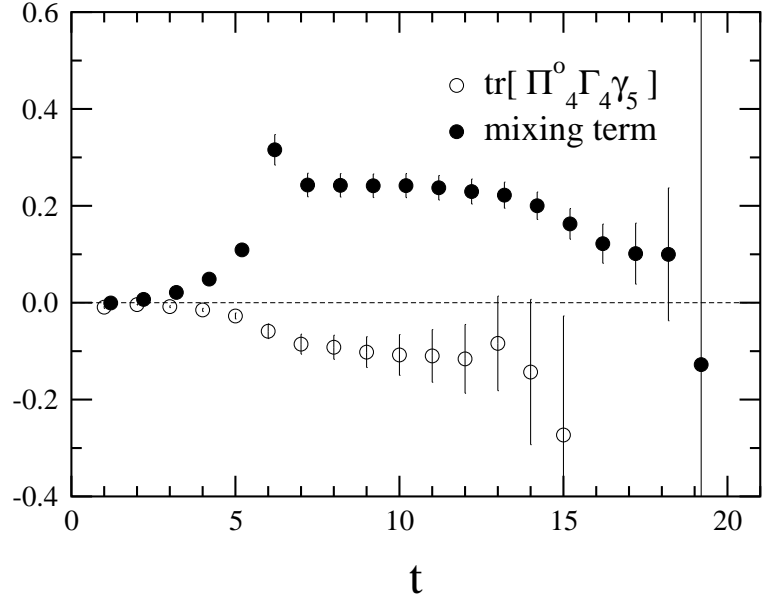}
\vskip 10mm 
\includegraphics[width=100mm, angle=0]
{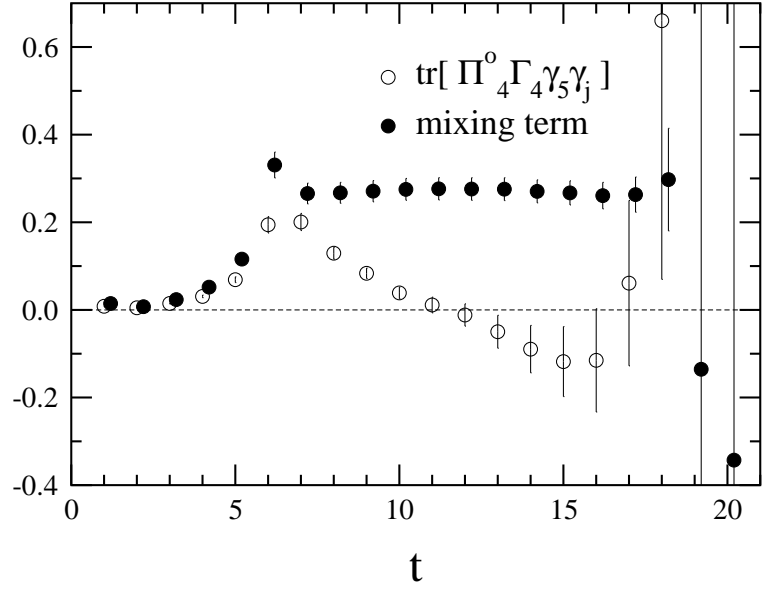}
\Ec
\caption{Parity-odd part of the form factors for the neutron.
 Top: $\tr[\Pi_4^o(q,t,\tau)\Gamma_{4}\gm_5]$ (open)
and the mixing term (filled) in eq.(\ref{f3_proj1}).
Bottom: $\tr[\Pi_4^o(q,t,\tau)\Gamma_{4}\gm_5\gm_i]$ (open)
and the mixing term (filled) in eq.(\ref{f3_proj2}).
}
\label{f3_split}
\end{figure}

\begin{figure}[h]
\Bc
\includegraphics[width=100mm, angle=0]
{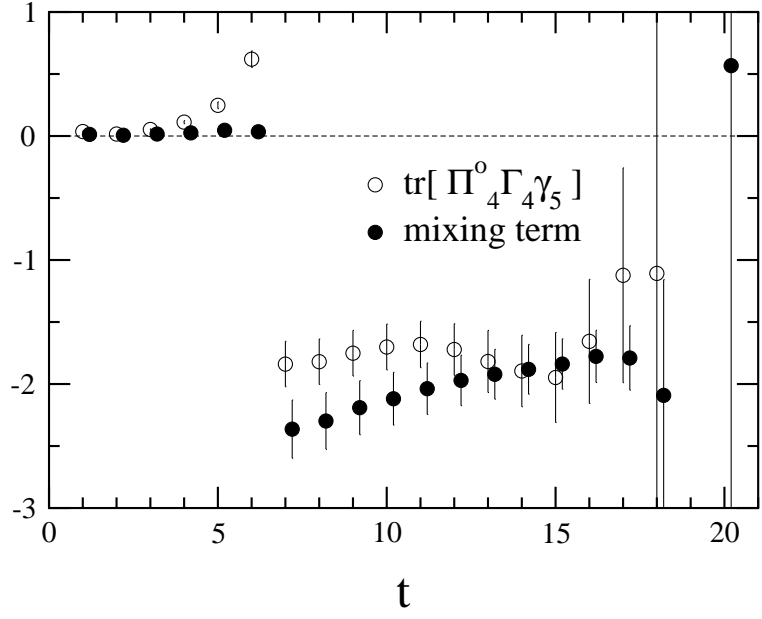}
\vskip 10mm 
\includegraphics[width=100mm, angle=0]
{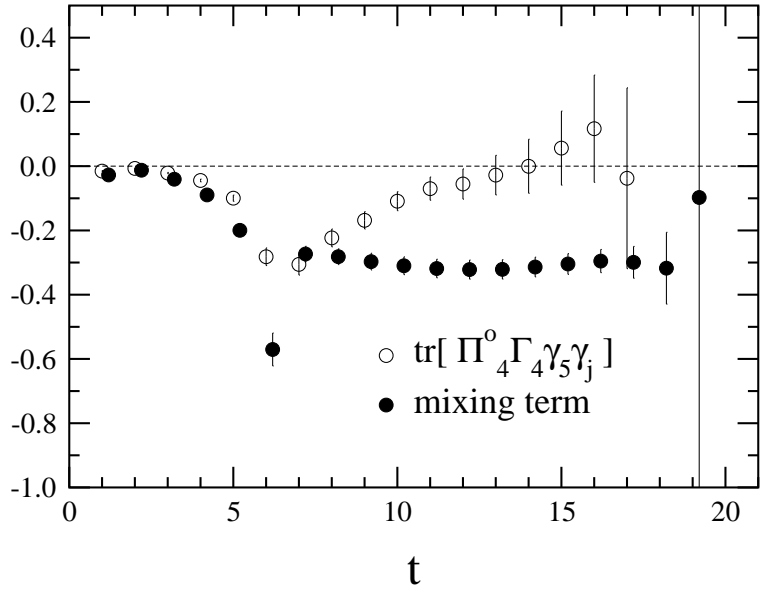}
\Ec
\caption{Same qunatities for the proton, as in the previous figure.
}
\label{f3_split2}
\end{figure}

We now consider the main target of our calculation: the parity-odd part
of 
the form factors.
Substituting $G_{NJ_\mu N}^Q(q;t,\tau)$, $F_1(q^2)$, $F_2(q^2)$ 
and $f_N^1$ into 
eqs.(\ref{f3_proj1}) and (\ref{f3_proj2}), we extract the form factor 
$F_3(q^2)$ in two different ways. 
We first plot 
$\tr[\Pi_4^o(q,t,\tau)\Gamma_{4}\gm_5]$ and
$\tr[\Pi_4^o(q,t,\tau)\Gamma_{4}\gm_5\gm_i]$ in Figs.~\ref{f3_split}
for the nucleon and in \ref{f3_split2} for the proton, together with the
second terms in the right hand side of 
eqs.(\ref{f3_proj1}) and (\ref{f3_proj2}),
composed of $F_1$, $F_2$ and $f_N^1$.
Comparing the two extractions, we find that
$\tr[\Pi_4^o(q,t,\tau)\Gamma_{4}\gm_5]$
and $\tr[\Pi_4^o(q,t,\tau)\Gamma_{4}\gm_5\gm_i]$ shown by open symbols
do not agree 
with each other.
In particular, for the case of the proton, 
$\tr[\Pi_4^o(q,t,\tau)\Gamma_{4}\gm_5]$
is very large, while $\tr[\Pi_4^o(q,t,\tau)\Gamma_{4}\gm_5\gm_i]$
is almost zero.

We note that 
$\tr[\Pi_4^o(q,t,\tau)\Gamma_{4}\gm_5]$ and
$\tr[\Pi_4^o(q,t,\tau)\Gamma_{4}\gm_5\gm_i]$ are almost
zero for the neutron is consistent with the previous result\cite{Blum},
in which the contribution of the mixing term was not considered.

\begin{figure}[h]
\Bc
\includegraphics[width=100mm, angle=0] {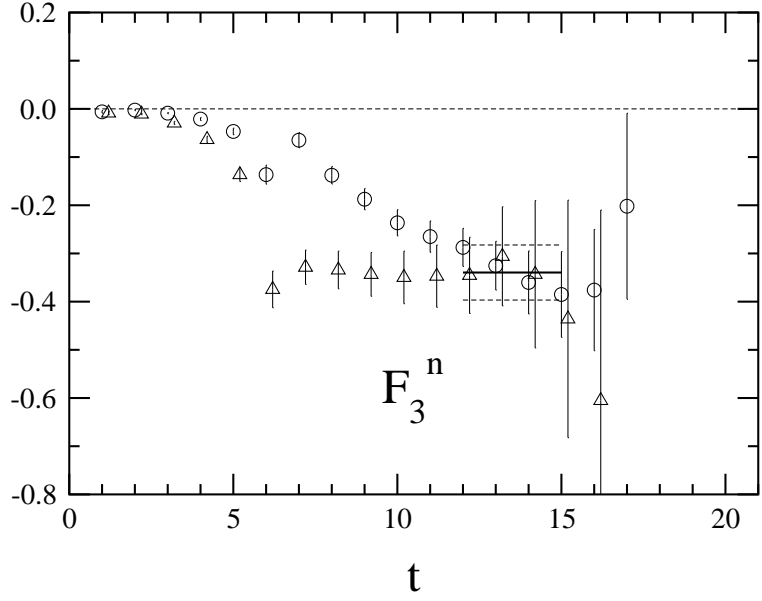}
\vskip 10mm 
\includegraphics[width=100mm, angle=0] {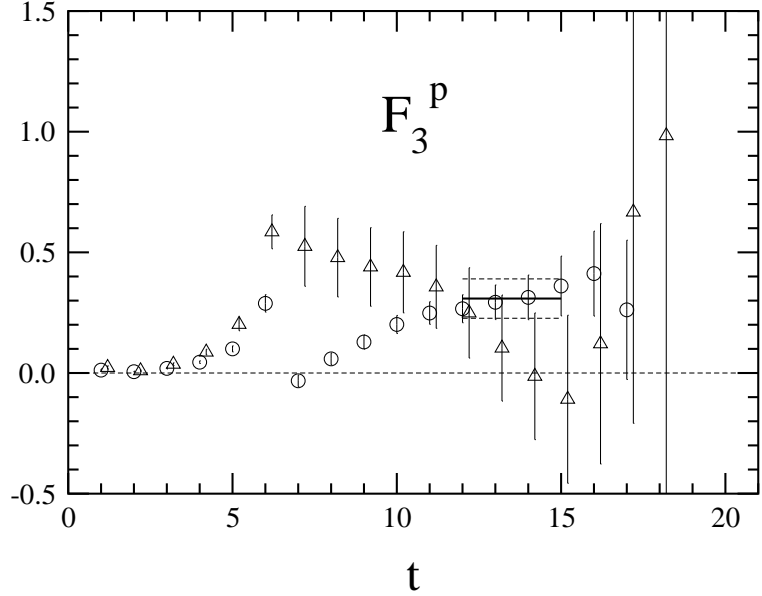}
\Ec
\caption{The parity odd form factor $F_3$  
for neutron (top) and proton (bottom).
Circles represent the result with 
$\Gamma_4\gm_5$ projection of eq.(\ref{f3_proj1}), 
while triangles represent the one with 
$\Gamma_4\gm_5\gm_j$ projection 
averaged over $j=1,2,3$ in eq.(\ref{f3_proj2}).}
\label{f3}
\end{figure}

Figure~\ref{f3} shows the result after subtracting the second terms from
$\tr[\Pi_4^o(q,t,\tau)\Gamma_{4}\gm_5]$ (circles) and
$\tr[\Pi_4^o(q,t,\tau)\Gamma_{4}\gm_5\gm_i]$ (triangles) for the neutron
(top panel) 
and the proton (bottom panel). 
It is gratifying to observe that the two ways of projections give
values of $F_3$ consistent with each other.
In particular, the significant difference between 
the two projections for the proton 
almost vanishes after removing the second terms.

{}Making a constant fit to the result of eq.(\ref{f3_proj2}) 
over $12\le t\le 15$,
we obtain 
\BE
  \frac{1}{2m_N}F_3(q^2\simeq 0.58\,\textrm{GeV}^2) = \LL\{
\begin{array}{rc}
  -0.024(5) \,\,e\cdot\textrm{fm}& (\textrm{neutron})\\
   0.021(6) \,\,e\cdot\textrm{fm}& (\textrm{proton})
\end{array}
\RR.\EE
with $a^{-1}=1.875(56)$GeV\cite{BK_CP-PACS}.
Note that the opposite sign between neutron and proton is consistent
with the estimate of Ref.\cite{ChPT}, 
where  the baryon EDM is proportional to its magnetic moment.  
It should be remembered that our values are obtained at a non-zero
$q^2$.
With the same caution in mind, our result may be compared with
other estimates  of ${F_3(0)}/{2m_N}$:
$3.6\times 10^{-3}\,e\cdot$fm\cite{Crewther},
$7.5\pm3.2\times 10^{-3}\,e\cdot$fm\cite{ChPT} and  
$3.9\times 10^{-3}\,e\cdot$fm\cite{Vecchia}.


\section{Conclusion and discussion}
\label{sec:conclusion}
In this paper we have carried out a feasibility study for the lattice
QCD 
calculation of the neutron
electric dipole moment in the presence of the $\theta$ term.
We took the strategy to extract the nucleon EDM from the nucleon 
electromagnetic form factor at small $\theta$.  
We explained  how one could extract 
the parity-odd part of the form factors, which becomes EDM at zero
momentum transfer,
from the relevant correlation functions for small $\theta$.
We have pointed out that 
the contribution of the parity-even form factors multiplied by
the parity-odd part of the nucleon two-point function 
has to be subtracted from the parity odd three-point 
function. 
We have derived the formula to carry out this subtraction, 
which is one of the main results.
 
In the second half, we applied our formula to an actual lattice QCD 
calculation, 
employing the domain-wall quark action with the RG improved gauge action
in quenched QCD.
We used a $16^3\times 32\times 16$ lattice at $a^{-1}=1.875(56)$GeV 
with the quark mass of $m_f=0.03$, which corresponds to $m_\pi/m_\rho =
0.629(8)$.
In order to obtain a sufficient sampling of topological charges,
we accumulated 730 configurations, which enabled us to extract the
parity-odd 
part both in the nucleon two- and three-point functions.
We have shown that two different $\gamma$ matrix projections yield
values
for $F_3$ consistent with each other, if the subtractions 
mentioned above are made. 

Our lattice calculation gives 
$F_3(q^2\simeq 0.58\,\textrm{GeV}^2)/(2m_N) =-0.024(5) \,\,e\cdot\textrm{fm}$ for neutron, 
$F_3(q^2\simeq 0.58\,\textrm{GeV}^2)/(2m_N) =0.021(6) \,\,e\cdot\textrm{fm}$ for proton 
at $m_fa=0.03$.  
In order to obtain the physical value of  NEDM,  we have to perform
various extrapolations: $q^2\rightarrow 0$ extrapolation, the chiral
extrapolation
($m_f\rightarrow m_{\rm phys}$) and the continuum extrapolation
($a\rightarrow 0$).
In addition, we have to remove the quenching error in NEDM,
since NEDM vanishes at zero quark mass in QCD 
but it does not in quenched QCD.
While it would be possible to remove each of these systematic errors 
step by step using the method proposed in 
this paper, we are now looking for better alternatives with which
we can remove them more easily.

\section*{Acknowledgments}
S. A. would like to thank Prof. A. Kronfeld for useful discussion, which
leads
to the simpler derivation of eq.(\ref{eq:2-pt_in_theta}). 
We thank Prof. M. Fukugita for his valuable comments on the manuscript.
This work is supported in part by Grants-in-Aid of the Ministry of
Education
(Nos.
12640253, 
13640259, 
13640260, 
14046202, 
15204015, 
15540251, 
15540279, 
15740134, 
15740165, 
16028201, 
16540228, 
17540249  
). The numerical simulations have been carried out on the parallel computer CP-PACS.

\appendix
\section{$N_+$ to $N_-$ form factor}\label{app1}
\begin{figure}[h]
\Bc
\includegraphics[width=100mm, angle=0] {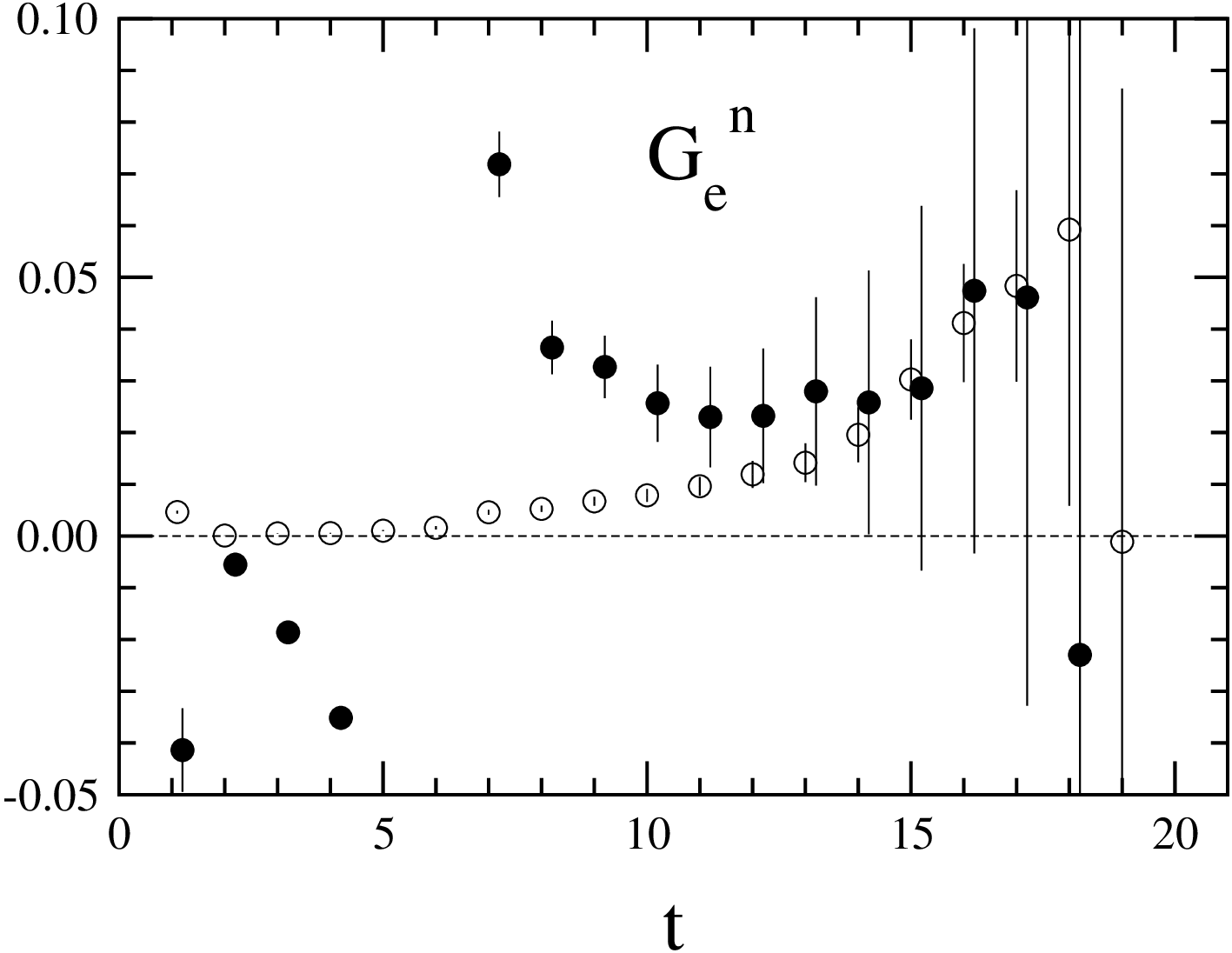}
\Ec
\caption{Solid circles represent the electric form factor $G_e^n(q^2)$
for 
neutron in different analysis where the effect of $N_-$ is taken into
account,
together with the previous result(open) in Fig.\ref{Ge}.}
\label{ge_nodd}
\end{figure}

If we include the electromagnetic transistion from $N_+$ to $N_-$, 
eqs.(\ref{elec_mom}) and (\ref{magne_mom}) are modified as
\BA
\tr[\Pi^e_4(q;t,\tau)\Gamma_4]
&=& \frac{E_N+m_N}{E_N}
    \LL(F_1(q^2) + F_2(q^2)\frac{q^2}{4m^2_N}\RR)\nn
&+& \bigg[
    \frac{\vec{q}^2}{E_{N_-}(m_N + m_{N_-})}F_3^{N_+\rightarrow
N_-}(q^2)\nn
&+& \frac{\tilde
q^2(E_{N_-}-m_{N_-})(1-\Lambda)}{E_{N_-}}F_A^{N_+\rightarrow N_-}(\tilde
q^2)
  \bigg] \Delta_{+-}(t,\tau), \label{fodd1}\\
\tr[\Pi^e_i(q;t,\tau)i\Gamma_4\gm_5\gm_j]
&=& \eps_{ijk}\frac{q_k}{E_N}\LL(F_1(q^2) + F_2(q^2)\RR)\nn
&+& \eps_{ijk}\bigg[
 -  \frac{q_k(m_N-m_{N_-})}{E_{N_-}(m_N+m_{N_-})}F_3^{N_+\rightarrow
N_-}(\tilde q^2)\nn
&+& \frac{q_k\tilde q^2}{E_{N_-}}F_A^{N_+\rightarrow N_-}(\tilde q^2)
   \bigg] \Delta_{+-}(t,\tau),\label{fodd2}
\EA
where 
\BE
  \Lambda = \frac{(m_N +
m_{N_-})(E_{N_-}-m_N)}{-(E_{N_-}-m_N)^2+\vec{q}^2},\quad
  \Delta_{+-}(t,\tau) =
\frac{(Z_-)^*Z_+}{|Z_+|^2}e^{-(E_{N_-}-E_N)(t-\tau)}
\EE
and $m_{N_-}$ is the parity-odd nucleon mass and 
$F_3^{N_+\rightarrow N_-}(\tilde q^2)$ and $F_A^{N_+\rightarrow
N_-}(\tilde q^2)$ 
are the $N_+$ to $N_-$ transition form factors with 
$\tilde q^2 = -(E_{N_-}-m_N)^2 + \vec{q}^2$. 
$Z_{N_-}(p)$ is the parity-odd nucleon amplitude which can be  
determined from the $N_-$ propagator.
In order to obtain these form factors we have to consider 
additional projection formula as follows:
\BA
\tr[\Pi^e_4(q;t,\tau)i\Gamma_4\gamma_j]
&=&  \frac{q_j}{E_N}\LL(F_1(q^2) + \frac{q^2}{2m_N}F_2(q^2)\RR) \nn
&+& \bigg[
    \frac{q_j(E_{N_-}+m_{N_-})}{E_{N_-}(m_N+m_{N_-})}F_3^{N_+\rightarrow
N_-}(\tilde q^2)\nn
&+& \frac{q_j\tilde q^2}{E_{N_-}}(1-\Lambda)F_A^{N_+\rightarrow
N_-}(\tilde q^2)
  \bigg] \Delta_{+-}(t,\tau),\label{fodd3}\\
\tr[\Pi^e_i(q;t,\tau)i\Gamma_4]
&=&  \frac{q_i}{E_N}F_1(q^2) - \frac{q_i(E_N-m_N)}{2m_NE_N}F_2(q^2)\nn
&+& \bigg[ 
    \frac{q_i(E_{N_-}-m_N)}{(m_N+m_{N_-})E_{N_-}}F_3^{N_+\rightarrow
N_-}(\tilde q^2)\nn
&-&  \frac{q_i\tilde
q^2(E_{N_-}-m_N+(-E_{N_-}+m_{N_-})\Lambda)}{E_{N_-}(E_{N_-}-m_N)}
    F_A^{N_+\rightarrow N_-}(\tilde q^2)
    \bigg] \Delta_{+-}(t,\tau),\nn\label{fodd4}\\
\tr[\Pi^e_i(q;t,\tau)\Gamma_4\gamma_j]
&=& \frac{-E_N+m_N}{E_N}\delta_{ij}F_1(q^2)
 + 
\frac{(E_N-m_N)^2\delta_{ij}-\vec{q}^2\delta_{ij}+q_iq_j}{2E_Nm_N}F_2(q^2)\nn
&-& \Bigg[   
\frac{(E_{N_-}+m_{N_-})(E_{N_-}-m_N)\delta_{ij}-\vec{q}^2\delta_{ij}+q_iq_j}{(m_N+m_{N_-})E_{N_-}}
    F_3^{N_+\rightarrow N_-}(\tilde q^2)\nn
&+&  \frac{\tilde
q^2}{E_{N_-}}\LL(-(E_{N_-}+m_{N_-})\delta_{ij}+\frac{\Lambda
q_iq_j}{E_{N_-}-m_N}\RR)
    F_A^{N_+\rightarrow N_-}(\tilde
q^2)\Bigg]\Delta_{+-}(t,\tau).\nn\label{fodd5}
\EA
Employing four equations out of the above five formula 
we can determine  
$F_1,\,F_2,\,F_3^{N_+\rightarrow N_-}$ and $F_A^{N_+\rightarrow N_-}$. 
The form factors $F_1$ and $F_2$ plotted 
in Figs.~\ref{ge_nodd} and \ref{f12_nodd}.
are obtained with the use of
eqs.(\ref{fodd3}), (\ref{fodd4}) and (\ref{fodd5})($i=j$ and $i\not=
j$).

\begin{figure}[h]
\Bc
\includegraphics[width=100mm, angle=0] {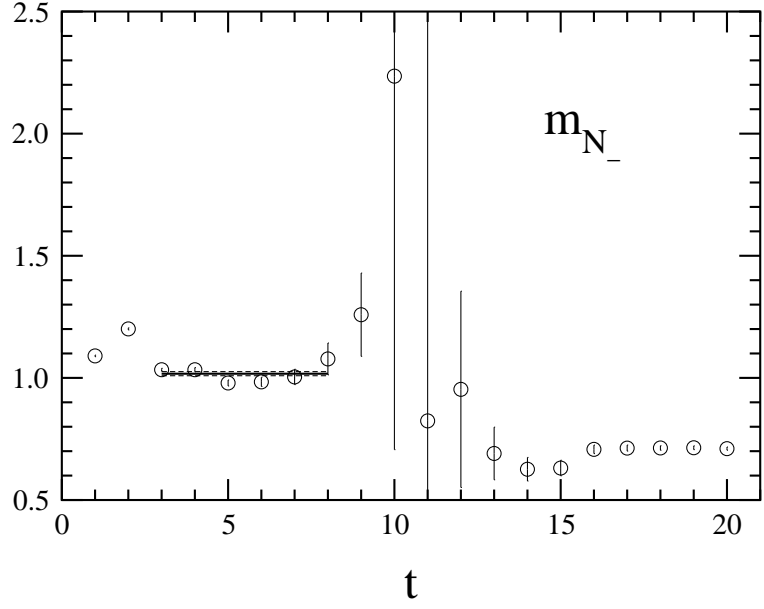}
\vskip 10mm 
\includegraphics[width=100mm, angle=0] {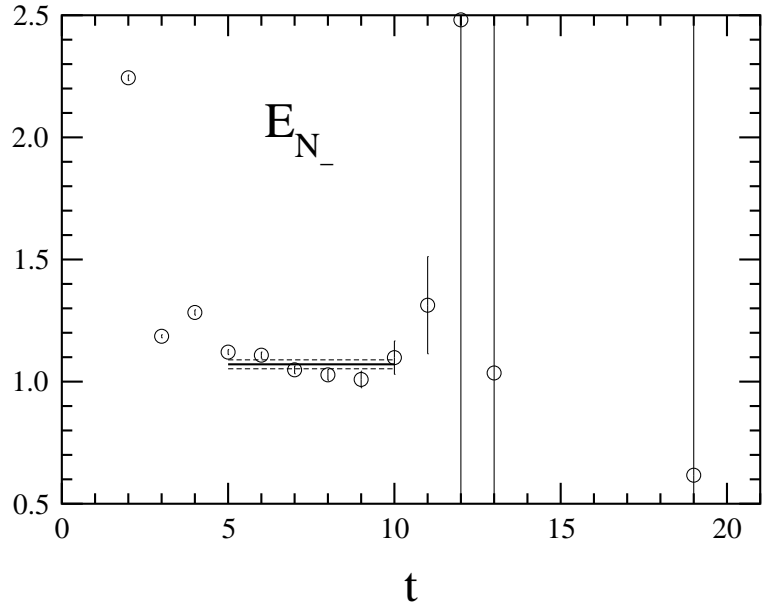}
\Ec
\caption{Effective mass plot for the parity-odd nucleon for 
$|{\vec p}|=0$ (top) and $|{\vec p}|=1$ (bottom).
Horizontal lines denote the central value (solid) and the errors
(dotted)
of the global fit.}
\label{Nem_odd}
\end{figure}

In Fig.~\ref{Nem_odd} we show the effective mass plot 
for the parity-odd nucleon.
As for $\vert {\vec p}\vert = 1$ we have subtracted the $N_+$
contribution
from $\tr [ G_{NN}(t,p)\bar\Gamma_4 ]$
with $\bar\Gamma_4 =\frac{1-\gamma_4}{2}$, in order to obtain the energy 
$E_{N_-}$ for the parity-odd state.
The global fit with $ 4\le t \le 9$ for $m_{N_-}$ and $ 6\le t \le 11$
for $E_{N_-}$ 
gives
\BE
 m_{N_-}a=1.017(8), E_{N_-}a=1.056(18),
\EE
where $E_{N_-}a$ is close to the value expected from the relativistic
dispersion relation $\sqrt{(m_{N_-}a)^2 +(\pi/8)^2}=1.090(8)$.

\vskip 5cm 
\newpage



\begin{figure}[h]
\Bc
\includegraphics[width=100mm, angle=0] {Fig/f1n.leading.eps}
\vskip 10mm 
\includegraphics[width=100mm, angle=0] {Fig/f1p.leading.eps}
\Ec
\caption{Form factor  $F_1(q^2)$ for neutron (top) and proton (bottom).
Horizontal lines represent the constant fit.}
\label{f1}
\end{figure}

\begin{figure}[h]
\Bc
\vskip 10mm 
\includegraphics[width=100mm, angle=0] {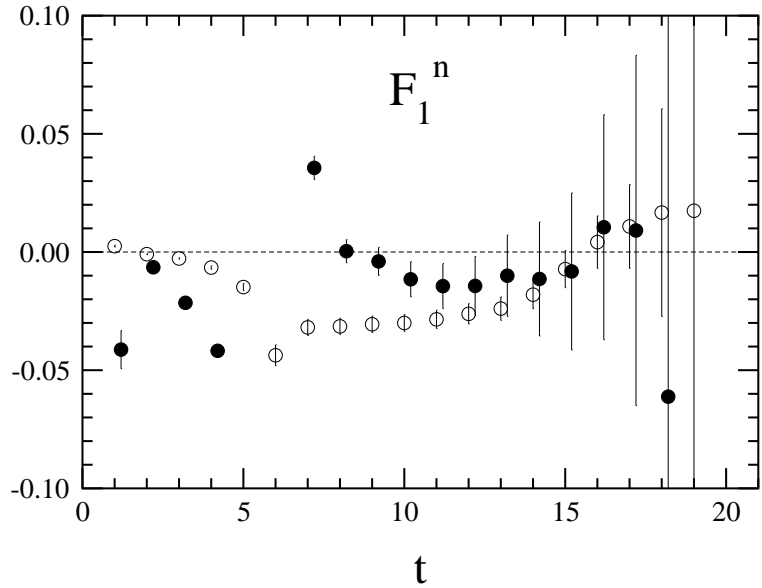}
\Ec
\caption{Solid circles represent the form factor $F_1^n(q^2)$ for 
neutron
in different analysis where the effect of $N_-$ is taken into account,
together with the previous result(open) in Fig.\ref{f1}.}
\label{f12_nodd}
\end{figure}


\begin{thebibliography}{90}
\bibitem{Harris}
P. G. Harris et al., Phys. Rev. Lett. {\bf 82}, 904 (1999). 
\bibitem{Dmitriev}
V. F. Dmitriev, R. A. Sen'kov, Phys. Rev. Lett. {\bf 91}, 212303 (2003).
\bibitem{Romalis}
M. V. Romalis, W. C. Griffith, J. P. Jacobs, E. N. Fortson, Phys. Rev.
Lett. {\bf 86}, 2505 (2001).
\bibitem{Crewther}
R. J. Crewther, P. Di Vecchia, G. Veneziano, E. Witten, 
Phys. Lett. {\bf B88}, 123 (1979); erratum, {\it ibid.} {\bf B91}, 487
(1980). 
\bibitem{Vecchia}
P. Di Vecchia, Acta Phys. Austriaca Suppl. {\bf 22}, 341 (1980).
\bibitem{Peccei}
R. D. Peccei, H. R. Quinn, Phys. Rev. Lett. {\bf 38}, 1440 (1977); 
R. D. Peccei, Adv. Ser. Direct. High Energy Phys. 3, 503-551 (1989). 
\bibitem{QCDsum}
M. Pospelov, A. Ritz, Nucl. Phys. {\bf B558}, 243 (1999); Nucl. Phys.
{\bf B573}, 177 (2000); 
Chuan-Tsung Chan, E. M. Henley, T. Meissner, hep-ph/9905317; 
M. Pospelov, A. Ritz, Phys. Rev. Lett. {\bf 27}, 2526 (1999); Phys. Rev.
{\bf D63}, 073015 (2001).
\bibitem{ChPT}
B. Borasoy, Phys. Rev. {\bf D61}, 114017 (2000). 
\bibitem{Aoki1}
S. Aoki, A. Gocksch, Phys. Rev. Lett. {\bf 63}, 1125 (1989); erratum,
{\it ibid.} {\bf 65}, 1172 (1990).
\bibitem{Aoki2}
S. Aoki, A. Gocksch, A. V. Manohar, S. R. Sharpe, Phys. Rev. Lett. {\bf
65}, 1092 (1990).
\bibitem{Guadagnoli}
D. Guadagnoli, V. Lubicz, G. Martinelli, S. Simula, JHEP {\bf 0304},019
(2003).
\bibitem{Faccioli}
P. Faccioli, D. Guadagnoli, S. Simula, Phys. Rev. {\bf D70}, 074017
(2004).
\bibitem{Blum}
F. Berruto, T. Blum, K. Orginos, A. Soni, Nucl. Phys. Proc. Suppl. 140,
411 (2005).
\bibitem{Aoki3}
S. Aoki, T. Hatsuda, Phys. Rev. {\bf D45}, 2427 (1992).
\bibitem{DW}
D. B. Kaplan, Phys. Lett. {\bf B288}, 342 (1992);
Y. Shamir, Nucl. Phys. {\bf B406}, 90 (1993); 
V.~Furman, Y.~Shamir, Nucl. Phys. {\bf B439}, 54 (1995).
\bibitem{overlap}
R. Narayanan, H. Neuberger, Nucl. Phys. {\bf B443}, 305 (1995); 
H. Neuberger, Phys. Lett. {\bf B417}, 141 (1998); 
H. Neuberger, Phys. Lett. {\bf B427}, 353 (1998); 
H. Neuberger, Phys. Rev. {\bf D57}, 5417 (1998).
\bibitem{CP-PACS1}
CP-PACS Collaboration, A. Ali Khan $et$ $al$., Phys. Rev. {\bf D63},
114504 (2001).
\bibitem{iwasaki}
Y. Iwasaki, Nucl. Phys. {\bf B258}, 148 (1985).
\bibitem{CP-PACS2}
CP-PACS Collaboration, A. Ali Khan $et$ $al$., Phys. Rev. {\bf D64},
114501 (2001).
\bibitem{impQ}
P. Weisz, Nucl. Phys. {\bf B212}, 1 (1983); 
P. Weisz, R. Wohlert, Nucl. Phys. {\bf B236}, 397 (1984); erratum, {\it
ibid.} {\bf B247}, 544 (1984);
M. L\"{u}scher, P. Weisz, Commun. Math. Phys. {\bf 97}, 59 (1985);
erratum, {\it ibid.} {\bf 98}, 
433 (1985).
\bibitem{BK_CP-PACS}
CP-PACS Collaboration, A. Ali Khan $et$ $al$., Phys. Rev. {\bf D64},
114506 (2001).
\bibitem{pdg2004}
S. Eidelman {\it et al.}, Phys. Lett. {\bf B592}, 1 (2004). 
\bibitem{QCDSF}
M. G\"{o}ckeler $et$ $al$., Phys. Rev. {\bf D71}, 034508 (2005).
\bibitem{Price}
L. E. Price, $et$ $al$., Phys. Rev. {\bf D4}, 45 (1971).
\bibitem{Gao}
H.-Y. Gao, Int. J. Mod. Phys. {\bf E12}, 1 (2003).
\bibitem{Xu}
W. Xu, $et$ $al$., Phys. Rev. {\bf C67}, R012201 (2003).
\bibitem{Zhu}
H.Zhu, $et$ $al$., Phys. Rev. Lett. {\bf 87}, 081801 (2001).
\bibitem{Tang}
A. Tang, W. Wilcox, R. Lewis, Phys. Rev. {\bf D68}, 094503 (2003).
\end{thebibliography}
\end{document}